\documentclass[aps,superscriptaddress,floats,floatfix,eqsecnum,nofootinbib]{revtex4}%
\usepackage{epsfig,graphicx,amssymb,amsbsy,bm,amsmath,rotating}
\usepackage{hyperref}
\usepackage{psfrag}

\newcommand{\be}{\begin{equation}}
\newcommand{\ee}{\end{equation}}
\newcommand{\bea}{\begin{eqnarray}}
\newcommand{\eea}{\end{eqnarray}}
\begin{document}
\title{CMB quadrupole suppression: II. The early fast roll stage.}
\author{D. Boyanovsky}
\email{boyan@pitt.edu} \affiliation{Department of Physics and
Astronomy, University of Pittsburgh, Pittsburgh, Pennsylvania 15260,
USA} \affiliation{Observatoire de Paris, LERMA. Laboratoire
Associ\'e au CNRS UMR 8112.
 \\61, Avenue de l'Observatoire, 75014 Paris, France.}
\affiliation{LPTHE, Universit\'e Pierre et Marie Curie (Paris VI) et
Denis Diderot (Paris VII), Laboratoire Associ\'e au CNRS UMR 7589,
Tour 24, 5\`eme. \'etage, 4, Place Jussieu, 75252 Paris, Cedex 05,
France}
\author{H. J. de Vega}
\email{devega@lpthe.jussieu.fr} \affiliation{LPTHE, Universit\'e
Pierre et Marie Curie (Paris VI) et Denis Diderot (Paris VII),
Laboratoire Associ\'e au CNRS UMR 7589, Tour 24, 5\`eme. \'etage, 4,
Place Jussieu, 75252 Paris, Cedex 05,
France}\affiliation{Observatoire de Paris, LERMA. Laboratoire
Associ\'e au CNRS UMR 8112.
 \\61, Avenue de l'Observatoire, 75014 Paris, France.}
\affiliation{Department of Physics and Astronomy, University of
Pittsburgh, Pittsburgh, Pennsylvania 15260, USA}
\author{N. G. Sanchez}
\email{Norma.Sanchez@obspm.fr} \affiliation{Observatoire de Paris,
LERMA. Laboratoire Associ\'e au CNRS UMR 8112.
 \\61, Avenue de l'Observatoire, 75014 Paris, France.}
\date{\today}
\begin{abstract}
Within the effective field theory of inflation, an initialization of
the classical dynamics of the inflaton with approximate
\emph{equipartition} between the kinetic and potential energy of the
inflaton leads to a brief {\bf fast roll stage} that precedes the
slow roll regime. The fast roll stage leads to an attractive
potential in the wave equations for the mode functions of curvature
and tensor perturbations. The evolution of the inflationary
perturbations is equivalent to the scattering by this potential and
a useful dictionary   between the scattering data and observables is
established. Implementing methods from scattering theory we prove
that this attractive potential leads to a {\bf suppression} of the
quadrupole moment for CMB and B-mode angular power spectra. The
scale of the potential is determined by the Hubble parameter during
slow roll.  Within the \emph{effective} field theory of inflation at
the grand unification (GUT) energy scale we find that if inflation
lasts a total number of efolds $ N_{tot} \sim 59 $, there is a $
10-20\% $ suppression of the CMB quadrupole and about $ 2-4\% $
suppression of the \emph{tensor} quadrupole. The suppression of
higher  multipoles is   smaller, falling off as $ 1/l^2 $. The
suppression is much smaller for $ N_{tot} > 59 $, therefore if the
observable suppression originates in the fast roll stage, there is
the {\bf upper bound} $ N_{tot} \sim 59 $.
\end{abstract}


\maketitle
\tableofcontents
\section{Introduction}

Scalar curvature and tensor (gravitational wave) quantum
fluctuations generated during the inflationary stage determine the
power spectrum of the anisotropies in the cosmic microwave
background (CMB) providing the seeds for large scale structure (LSS)
formation. Curvature and tensor fluctuations obey a wave equation,
and the choice of a particular solution entails a choice of initial
conditions\cite{kolb}-\cite{CMBgiova}. The power spectra of these
fluctuations depend in general on the initial conditions that define
the particular solutions. These   are usually chosen as
Bunch-Davies\cite{BD} initial conditions, which select positive
frequency modes asymptotically with respect to conformal time. The
quantum states in the Fock representation associated with these
initial conditions are known as Bunch-Davies states, the vacuum
state being invariant under the maximal symmetry group $ O(4,1) $ of
de Sitter space-time. In earlier studies alternative initial
conditions were also considered\cite{diferen}.  The requirement that
the energy momentum tensor be renormalizable constrains the UV
asymptotic behaviour of the Bogoliubov coefficients that encode
different initial conditions\cite{motto2}. The availability of high
precision cosmological data motivated a substantial effort to study
the effect of different initial conditions upon the angular power
spectrum of CMB anisotropies, focusing primarily in the high-l
region near the acoustic peaks\cite{otros}. However, the exhaustive
analysis of the three year WMAP data\cite{WMAP31}-\cite{WMAP33}
render much less statistical significance to possible effects on
small angular scales from alternative initial conditions.

Although there are no statistically significant departures from the
slow roll inflationary scenario at small angular scales ($ l\gtrsim
100 $), the third year WMAP data again confirms the surprinsingly
low quadrupole $ C_2 $ \cite{WMAP31}-\cite{WMAP33} and suggests that
it cannot be completely explained by galactic foreground
contamination. The low value of the quadrupole has been an
intriguing feature on large angular scales since first observed by
COBE/DMR \cite{cobe}, and confirmed by the WMAP data
\cite{WMAP31}-\cite{peiris}.

In a companion article \cite{I},  we
reported on our  study of  the effect of general initial conditions
on the power spectra of curvature and gravitational wave
perturbations. General initial conditions are related to the
Bunch-Davies initial conditions by a Bogoliubov transformation and
their effect on the power spectra is encoded in a \emph{transfer
function } $D(k)$ whose large wavevector behavior is constrained by
renormalizability and small backreaction\cite{I}. The rapid falloff of
$ D(k) \lesssim \mathcal{O}(1/k^2) $ for large $k$ entail   that
observable effects from initial conditions are more pronounced for
\emph{low multipoles}, namely in the region of the angular power
spectra corresponding to the Sachs-Wolfe plateau.

In ref. \cite{I} we formulate the problem of initial conditions
established at the beginning of slow roll, in terms of a scattering
by a potential in the wave equations for the mode functions of
curvature and tensor perturbations. Such potential is localized in
conformal time prior to slow roll and determines the initial
conditions for the mode functions. Implementing methods from
potential scattering allowed us to establish that such potential
yields a transfer function $D(k)$ that automatically satisfies the
stringent constraints from renormalizability and backreaction. The
results of this previous study reveal that an attractive potential
localized just prior to the onset of slow roll and with a scale
determined by the energy scale during slow roll inflation yield a
\emph{suppression } of the quadrupole for curvature perturbations
consistent with the data $\sim 10-20\%$ and predicts a small
quadrupole suppression for tensor perturbations.

In this article we discuss the \emph{origin  of this attractive
potential within the effective field theory of inflation}.
We show that such potential is a \emph{generic} feature of a brief
\emph{fast roll} stage that merges smoothly with   slow roll
inflation. This stage is a consequence of an initial condition
\emph{for the classical inflaton dynamics} in which the kinetic and
potential energy of the inflaton are of the same order, namely, the
energy scale of slow roll inflation. During the early fast roll
stage the inflaton evolves rapidly during a brief period,  but slows
down by the cosmological expansion settling in  the slow roll stage
in which the kinetic energy of the inflaton is much smaller than its
potential energy. The scale of the attractive potential is
determined by the energy scale during the slow roll stage, which in
the effective field theory description\cite{1sN,clar} is of the
order of  the grand unification scale, $ M \sim 10^{16}$GeV, well
below the Planck scale $ M_{Pl} \sim 10^{19}$GeV, and no other
energy scales are involved. Hence, we emphasize that there is {\bf
no need} to advocate transplanckian physics in this context.

\bigskip

\textbf{Brief summary of results :}

\medskip

In this article we combine the {\bf dynamical} origin of the
potential within the effective field theory of inflation, with the
results obtained in ref.\cite{I} and show that the early fast roll
stage leads to a suppression of the CMB quadrupole .

Our main results are the following:

\begin{itemize}

\item{Within the effective field theory  of
inflation with the same inflaton potentials that fulfill the slow
roll conditions, we find that an initial state of the inflaton with
almost \emph{equipartition} between kinetic and potential inflaton
energies yields an \emph{attractive} potential for the mode
functions of the fluctuations. This potential emerges from a brief
stage in which the inflaton rolls fast, hence we call this the
{\bf fast roll} stage. This early stage only lasts approximately
one e-fold   and merges smoothly with the slow roll stage. This fast
roll stage prior to slow roll is a \emph{generic} feature of an
initial condition for cosmological dynamics in which there is an
approximate equipartition between the kinetic and potential energy
of the inflaton. The initial conditions for the fluctuations
\emph{prior} to the fast roll stage are chosen to be the usual
Bunch-Davies conditions. However, the potential that results from
the fast roll   dynamics of the inflaton lead to non-Bunch Davis
conditions for the curvature and tensor perturbations at the
beginning of the slow roll stage. The Bogoliubov coefficients and
transfer function $D(k)$ automatically satisfy the constraints from
renormalizability and small backreaction.}

\item{We have investigated a large variety of inflationary models with initial inflaton dynamics
featuring an approximate equipartition between inflaton kinetic and
potential energies. This study leads us to  conclude quite generally
that the scale of the potential during   fast roll
  is completely determined by the
Hubble scale during the subsequent slow roll stage. The effect of
this potential during the {\bf fast roll} evolution of the scale
factor leads to modifications of the primordial power spectrum. This
potential is \emph{attractive} both for curvature and tensor
fluctuations, and leads to a {\bf suppression} of their primordial
power spectra on large scales.  }

\item{From a comprehensive numerical study of different inflationary scenarios
within the effective field theory approach, we find a $ 10-20\% $
\emph{suppression} of the CMB quadrupole and about a $ 2-4\% $
suppression of the B-mode quadrupole (tensor fluctuations). This CMB
quadrupole corresponds to the wavevector $ k_Q $ whose physical
wavelength is of the order of the Hubble radius \emph{today} and
exits the horizon during slow roll inflation just $ 1-2 $ e-folds
after the brief fast roll stage. The suppression on higher
$l$-multipoles reduce considerably following a  $ 1/l^2 $ law. }

\item{The attractive potential resulting from the fast roll stage   accounts for the observed suppression of the CMB quadrupole
\emph{if} the wavevector $k_Q$ whose wavelength corresponds to the
Hubble radius \emph{today} exits 2-3 e-folds after the end of the
fast roll stage, which lasts $\approx 1$ e-fold. The quadrupole
corresponds to the wavevector $k_Q$ that exits the horizon $ N_Q =
55 $ efolds before the end of inflation, hence  our results
successfully explain  the CMB quadrupole suppression within the
effective field theory if inflation lasts at most $ N_{tot} \lesssim
N_Q + 4 = 59 $ efolds. This result establishes an  {\bf upper bound}
to the number of efolds during inflation.}

\end{itemize}

\section{Initial Conditions of Inflationary Fluctuations
from the Scattering by a Potential} \label{condinfr}

In the companion article \cite{I}  we have systematically analyzed
the consequences of generic initial conditions different from
Bunch-Davies, under the conditions that these  are UV allowed and
yield   small backreaction effects. Here we address  the
\emph{origin} of these initial conditions,  beginning  by gathering
relevant ingredients from\cite{I}.

As shown in\cite{I}   in a cosmological space-time geometry
$$
ds^2 = dt^2-a^2(t)(d\vec x)^2 = C^2(\eta)[d\eta^2 - (d\vec x)^2] \; ,
$$
where $ t $ and $ \eta $ stand for cosmic and conformal time
respectively, the wave equations for the mode functions of gaussian
curvature and tensor perturbations are of the form of the
Schr\"odinger equation in one dimension \be \label{sceq}
\Big[\frac{d^2}{d\eta^2}+k^2-W(\eta)\Big]S(k;\eta) =0 \ee with $
\eta $ the coordinate, $ k^2 $ the energy and $ W(\eta) $ a
potential that depends on the coordinate $ \eta $. In the cases
under consideration \be W(\eta)  = \Bigg\{ \begin{array}{l}
W_{\mathcal{R}}(\eta) = z''/z  ~~\mathrm{for~curvature~perturbations \; ,}        \\ \\
W_T(\eta) = C''/ C  ~~\mathrm{for~tensor~perturbations \; .} \\
\end{array} \;   \label{defW}
\ee where prime stands for derivative with respect to the conformal
time and \be \label{za} z= a(t) \; \frac{\dot{\Phi} }{H} \; , \ee $
\dot \Phi  $ stands for the derivative of the inflaton field $ \Phi
$ with respect to the cosmic time $ t $.

\medskip

It is convenient to  explicitly separate the behavior of $ W(\eta) $
during the slow roll stage by writing
\be\label{defV}
W(\eta)=  \mathcal{V}(\eta) +\frac{\nu^2-\frac14}{\eta^2} \; ,
\ee
where
\be
\nu =  \Bigg\{ \begin{array}{l}
\nu_{\mathcal{R}}=\frac32+ 3 \, \epsilon_v -\eta_v +  {\cal O}\left(\frac1{N^2}\right)
~~\mathrm{for~curvature~perturbations}        \\ \\
\nu_T = \frac32+ \epsilon_v + {\cal O}\left(\frac1{N^2}\right)
 ~~\mathrm{for~tensor~perturbations} \; .  \\
\end{array} \label{defnu}
\ee Here $ \epsilon_v $ and $ \eta_v $ stand for the slow roll
parameters \be \epsilon_v = \frac{{\dot \Phi}^2}{2 \;  M^2_{Pl} \;
H^2}= \frac{M^2_{Pl}}2 \; \left[\frac{V'(\Phi)}{V(\Phi)}
\right]^2  +  {\cal O}\left(\frac1{N^2}\right) = {\cal
O}\left(\frac1{N}\right)   \quad , \quad \eta_v = M^2_{Pl} \;
\frac{V''(\Phi)}{V(\Phi)}  = {\cal O}\left(\frac1{N}\right) \; ,
\label{slowroll} 
\ee 
and $ N \sim 55 $ stands for the number of
efolds from horizon exit until the end of inflation \cite{1sN}.

The slow roll dynamics acts through the term $ [(\nu^2-1/4)/(\eta^2)] $
which is a \emph{repulsive}  centrifugal barrier.\\

We anticipate that the potential $ \mathcal{V}(\eta) $ is localized
in the fast roll stage \emph{prior} to   slow roll  (during which
cosmologically relevant modes cross out of the Hubble radius) where
$ \mathcal{V}(\eta) $ vanishes. In terms of the potential
$\mathcal{V}(\eta)$ the equations for the quantum fluctuations read,
\be 
\left[\frac{d^2}{d\eta^2}+k^2-\frac{\nu^2-\frac14}{\eta^2}-
\mathcal{V}(\eta)  \right]S(k;\eta) = 0 \;.\label{eqnpsr2} 
\ee
During the slow roll stage $ \mathcal{V}(\eta) = 0 $ and the mode
equations simplify to 
\be 
\left[\frac{d^2}{d\eta^2}+k^2 -
\frac{\nu^2 -\frac14}{\eta^2} \right]S(k,\eta) = 0 \; .
\label{geneq} 
\ee 
To leading order in slow roll, $ \nu $ is constant
and for general initial conditions the solution is, \be \label{genS}
S (k;\eta) = A (k)\,g_{\nu }(k;\eta) + B (k) \,f_{\nu }(k;\eta) \; ,
\ee where two linearly independent solutions of eq.(\ref{geneq})
are, \bea g_{\nu }(k;\eta) & = & \frac12 \; i^{\nu +\frac12} \;
\sqrt{-\pi \eta}\,H^{(1)}_{\nu }(-k\eta) \; , \label{gnu}\\
f_{\nu }(k;\eta) & = & [g_{\nu }(k;\eta)]^*\label{fnu}  \; ,
\eea
\noindent $ H^{(1)}_\nu(z) $ are Hankel functions. These
solutions are normalized so that their Wronskian is given by
\be
W[g_\nu(k;\eta),f_\nu(k;\eta)]=
g'_\nu(k;\eta) \; f_\nu(k;\eta)-g_\nu(k;\eta) \; f'_\nu(k;\eta) = -i \; .
\label{wronskian}
\ee
The mode functions and coefficients $ A(k), \; B(k) $ will feature a subscript index
$ {\mathcal{R}}, \; T $, for curvature or tensor perturbations, respectively.

For wavevectors deep inside the Hubble radius $ | k \, \eta | \gg 1
$, the mode functions have the Bunch-Davies asymptotic behavior
\be
g_{\nu}(k;\eta) \buildrel{\eta \to
-\infty}\over= \frac1{\sqrt{2 \, k}} \; e^{-ik\eta} \quad ,  \quad
f_{\nu}(k;\eta) \buildrel{\eta \to -\infty}\over=
\frac1{\sqrt{2k}} \; e^{ ik\eta} \; , \label{fnuasy}
\ee
and for $ \eta \to 0^- $, the mode functions behave as:
\be\label{geta0}
g_\nu(k;\eta)\buildrel{\eta \to 0^-}\over=
\frac{\Gamma(\nu)}{\sqrt{2 \, \pi \; k}} \; \left(\frac2{i \; k \;
\eta} \right)^{\nu - \frac12} \; .
\ee
The complex conjugate formula holds for $ f_{\nu}(k;\eta) $.

In particular, in the scale invariant case $ \nu=\frac32 $ which is
the leading order in the slow roll expansion, the mode functions eqs.(\ref{gnu})
simplify to
\be
g_{\frac32}(k;\eta) =
\frac{e^{-ik\eta}}{\sqrt{2k}}\left[1-\frac{i}{k\eta}\right]\,.\label{g32}
\ee
The mode equation (\ref{eqnpsr2}) can be written as an integral equation,
\be \label{solu}
S(k;\eta)= g_\nu(k;\eta) + i
\; g_\nu(k;\eta)\,\int^{\eta}_{-\infty}
 g^*_\nu(k;\eta') \; \mathcal{V}(\eta') \;  S(k;\eta') \; d\eta'-i  \;
g^*_\nu(k;\eta)\,\int^{\eta}_{-\infty}
 g_\nu(k;\eta') \; \mathcal{V}(\eta') \;  S(k;\eta') \; d\eta'\quad .
\ee
 This solution has the Bunch-Davies asymptotic condition
\be S(k;\eta \rightarrow -\infty) = \frac{e^{-ik\eta}}{\sqrt{2k}}
\,. \ee We formally consider here the conformal time starting at $
\eta = -\infty $. However, it is natural to consider that the
inflationary evolution of the universe starts at some negative value
$ \eta_i < {\bar \eta} $, where $ {\bar \eta} $ is  the conformal
time when fast roll ends and slow roll begins.

\medskip

Since $ \mathcal{V}(\eta) $ vanishes for  $ \eta > {\bar \eta} $,  the
mode functions $ S(k;\eta) $ can be written for  $ \eta > {\bar \eta} $
as linear combinations of the mode functions $ g_\nu(k;\eta) $ and $
g^*_\nu(k;\eta) $,
\be \label{solSR}
S(k;\eta) = A(k) \; g_\nu(k;\eta) + B(k) \; g^*_\nu(k;\eta)
\quad , \quad \eta > {\bar \eta} \quad ,
\ee
where the coefficients $ A(k) $ and $ B(k) $ can be read from eq.(\ref{solu}),
\bea
A(k) & = &  1+ i\int^{0}_{-\infty}
g^*_\nu(k;\eta) \; \mathcal{V}(\eta) \;  S(k;\eta) \;
d\eta\label{aofk} \cr \cr
 B(k) & = & -i \int^{0}_{-\infty}  g_\nu(k;\eta) \; \mathcal{V}(\eta)  \;
S(k;\eta) \;  d\eta\label{bofk}  \; .
\eea
The coefficients $ A(k) $ and $ B(k) $ are therefore {\bf calculated}
from the {\bf dynamics  before} slow roll [recall that $ \mathcal{V}(\eta)  = 0 $
for  $ \eta > {\bar \eta} $ during slow roll.]

\noindent The constancy of the Wronskian
$ W[g_\nu(\eta),g^*_\nu(\eta)]=-i $ and eq.(\ref{solSR}) imply the
constraint,
$$
|A(k)|^2-|B(k)|^2=1 \quad .
$$
This relation permits to represent the  coefficients $ A(k); \; B(k) $ as \cite{I}
\be
A(k) = \sqrt{1+N(k)} \; e^{i\theta_A(k)}~~;~~
B(k)=\sqrt{N(k)} \; e^{i\theta_{B}(k)} \label{bogonum} \; ,
\ee
\noindent where $ N(k), \; \theta_{A,{B}}(k) $ are real.

\medskip

Starting  with Bunch-Davies initial conditions for $ \eta \to
-\infty $, the action of the potential generates   a mixture of the
two linearly independent mode functions that result in the mode
functions eq.(\ref{solSR}) for $ \eta
> {\bar \eta} $ when the potential vanishes. This is clearly equivalent to starting the evolution
of the fluctuations at the \emph{beginning} of slow roll $ \eta =
{\bar \eta} $ with initial conditions defined by the Bogoliubov
coefficients $ A(k) $ and $ B(k) $ given by eq.(\ref{bofk}).

As shown in ref.\cite{I} the power spectrum of curvature and tensor
perturbations for the general fluctuations eq.(\ref{solSR}) takes
the form, 
\bea\label{curvapot} 
&& P_\mathcal{R}(k) \buildrel{\eta
\to  0^-}\over= \frac{k^3}{2 \; \pi^2} \; 
\Big|\frac{S_\mathcal{R}(k;\eta)}{z} \Big|^2 =
P^{sr}_{\mathcal{R}}(k)\Big[1+D_\mathcal{R}(k) \Big] \; , \cr \cr 
&& P_T(k)  \buildrel{\eta \to 0^-}\over= \frac{k^3}{2 \; \pi^2} \;
\Big|\frac{S_T(k;\eta)}{C(\eta)} \Big|^2
= P^{sr}_T(k)\Big[1+ D_T(k) \Big] \; . 
\eea
where $ D_\mathcal{R}(k) $ and $ D_T(k) $ are the transfer functions
for the initial conditions of curvature and tensor perturbations
introduced in ref.\cite{I}: 
\bea \label{DofkR} 
D_\mathcal{R}(k) &=&
2 \; | {B}_\mathcal{R}(k)|^2 -2 \; \mathrm{Re}\left[A_\mathcal{R}(k)
\; B^*_\mathcal{R}(k)\,i^{2\nu_R-3}\right] = 2 \; N_\mathcal{R}(k)-2
\; \sqrt{N_\mathcal{R}(k)[1+N_\mathcal{R}(k)]} \; \cos\left[
\theta_k^\mathcal{R} - \pi \left(\nu_R - \frac32 \right) \right]\cr \cr 
D_T(k) &=&2 \; | {B}_T(k)|^2 -2 \;
\mathrm{Re}\left[A_T(k)B^*_T(k)\,i^{2\nu_T-3}\right] = 2 \; N_T(k)-2
\;\sqrt{N_T(k)[1+N_T(k)]}\, \cos\left[ \theta_k^T - \pi \left(\nu_T
- \frac32 \right) \right] \; , 
\eea 
here $  \theta_k \equiv
\theta_{B}(k)-\theta_A(k) $. The standard slow roll power spectrum
is given by \cite{liddle,rmp}: \bea  \label{potBD}
P^{sr}_\mathcal{R}(k) &=& \left( \frac{k}{2 \, k_0}\right)^{n_s - 1}
\; \frac{\Gamma^2(\nu)}{\pi^3} \; \frac{H^2}{2 \; \epsilon_v  \;
M_{Pl}^2 }\equiv \mathcal{A}^2_\mathcal{R} \;
\left(\frac{k}{k_0}\right)^{n_s - 1} \; ,\cr \cr P^{sr}_T(k) &=&
\mathcal{A}^2_T \; \left(\frac{k}{k_0}\right)^{n_T} \;   \quad ,
\quad n_T= -2 \; \epsilon_v \quad , \quad
\frac{\mathcal{A}^2_T}{\mathcal{A}^2_\mathcal{R}} = r = 16 \;
\epsilon_v \; . \eea As shown in ref. \cite{I}, the relative change
in the $ C_l's $  for the  general fluctuations eq.(\ref{solSR})
with respect to the standard slow roll result is given by \be C_l
\equiv C^{sr}_l + \Delta C_l \quad , \quad \frac{\Delta C_l}{C_l} =
\frac{\int^\infty_0 D(\kappa\,x)~ f_l(x)\,dx}{\int^\infty_0
f_l(x)\,dx} \label{DelC} \; , \ee where $ x= k/\kappa $ and \be
\kappa \equiv a_0 \;  H_0/3.3   \; . \label{kappa} \ee $
D(\kappa\,x) $ is the transfer function of initial conditions for
the corresponding perturbation, \be f_l(x) = x^{n_s-2}[j_l(x)]^2 \,.
\label{fls} \ee and the $ j_l(x) $ are spherical Bessel functions
\cite{abra}. We derived in ref.\cite{I} an estimate of the
corrections, for the maximal asymptotic decay of the occupation
numbers \be \label{ocupa} N_k = N_\mu\,
\left(\frac{\mu}{k}\right)^{4+\delta}~~;~~0 < \delta \ll 1 \ee with
the result, \be \frac{\Delta C_l}{C_l} \approx - \frac43  \;
\sqrt{N_{\mu}} \;  \left(\frac{3.3\,\mu}{a_0 \; H_0} \right)^ 2 \,
\frac{\overline{\cos\theta}}{(l-1)(l+2)} \; . \label{Cl10} \ee where
we have taken $\nu =3/2$ and $\cos\theta_k \approx
\overline{\cos\theta}$ (see ref.\cite{I} for details). The $ \sim
1/l^2 $ behavior is a result of the $ 1/k^2 $ fall off of $ D(k) $,
a consequence of the renormalizability condition on the occupation
number. For the quadrupole, the relevant wave-vectors correspond to
$ x \sim 2 $, namely  $ k_Q \sim a_0 \; H_0 $. It is convenient to
write \be k_Q = a_{sr} \; H_{i} = a_0 \; H_0 \label{kH} \; , \ee
where $ a_{sr} $ and $ H_{i} $ are the scale factor and the Hubble
parameter during the slow roll stage of inflation when the
wavelength corresponding to \emph{today}'s Hubble radius exits the
horizon.

\section{The origin of the potential $ \mathcal{V}(\eta) $
: a \emph{fast roll} stage before slow roll inflation.}\label{fasr}

The mode functions of perturbations obey the general evolution
equation (\ref{sceq})
where $ W(\eta)  $ is given by eq.(\ref{defW}) and the slow roll part is explicitly
separated in eq.(\ref{defV}). A full expression for $ W(\eta) $ and
therefore for the potential $ \mathcal{V}(\eta) $ is obtained from the
Friedmann equation and the evolution equation of the inflaton
\bea\label{infa}
&&H^2 = \frac1{3 \; M^2_{PL}}\left[\frac12 \; {\dot \Phi}^2 +
V(\Phi)\right] \; , \\ \cr
&&\ddot\Phi + 3 \; H  \; \dot \Phi  + V'(\Phi) = 0  \; , \label{infa2}
\eea
The exact potential $ W(\eta) $ is obtained by using the
equations (\ref{infa})-(\ref{infa2}). For this purpose it is
convenient to introduce a dimensionless variable $ y^2 $ as
\be\label{y} 
y^2 \equiv \frac{{\dot \Phi}^2}{2 \;  M^2_{Pl} \; H^2}
=3 \left[ 1 - \frac {V(\Phi)}{3 \; M^2_{Pl}  \; H^2}\right] \quad ,
\quad 0\leq y^2 \leq 3 \; , 
\ee 
in terms of which the equations of
motion (\ref{infa}) and (\ref{infa2}) are written in the simple
form, 
\be \label{hpu} 
\dot\Phi =  \textrm{sign}(\dot{\Phi})\, M_{Pl}
\; H \; \sqrt2 \; |y| \quad  , \quad \frac{\dot H}{H^2} = - y^2 \; .
\ee 
In particular, during the slow roll stage: $ y^2 = \epsilon_v
$ [see eq.(\ref{slowroll})], but in general, in a stage in which the
slow roll approximation is not valid, the kinetic term of the
inflaton is not small. The slow roll parameters given by eqs.(\ref{slowroll})
are $ \epsilon_v \ll 1, \; \eta_v \ll 1 $ to correctly describe the
slow roll stage. But, besides the slow roll stage, in which $ y^2
\ll 1 $, there is a prior stage in which $ y^2 $ is \emph{not small}
but $ y^2 \sim \mathcal{O}(1) $: in this case the kinetic term of
the inflaton is of the same order as the potential $ V(\Phi) $. That
is, the initial energy of the inflaton is distributed between
kinetic and potential energy with approximate \emph{equipartition}.

\medskip

Thus, there are two distinct regimes determined by the dimensionless
variable $ y^2 $: (i) $ y^2 = {\cal O}\left(\frac1{N}\right) \ll 1 $
corresponds to the usual slow roll regime $ {\dot \Phi}^2 \ll V(\Phi) $;
(ii) in contrast, $ y^2 \gtrsim 1 $ in which $ \dot{\Phi}^2 \sim V(\Phi) $
describes a {\bf fast roll} regime. Inflation requires:
\be\label{ynfla}
\frac{\ddot a}{a} = H^2 \; (1 - y^2) >0 ~~~~, \quad
\ee
thus, the range of the variable $ y^2 $ for inflationary evolution is $ 0<y^2<1 $.

\subsection{Fast Roll Dynamics}

Notice that the same description of inflation (the same inflaton potential) gives rise to
the two diferent regimes: \emph{fast} roll {\bf and}  slow roll regimes. The dynamics in the
effective field theory of inflation giving rise to a fast roll stage followed by
the slow roll stage is simple: consider an initial condition on the
inflaton field and its first derivative that corresponds to an
initial value of $ y^2 \sim 1 $. The potential and kinetic energy of
the inflaton in this state are of the same order, this is the
beginning of the \emph{fast roll} stage. The strong friction term in
the equation of motion for the inflaton eq.(\ref{infa}) results in that
if initially $ \dot{\Phi} \neq 0 $ and large, the kinetic energy of
the inflaton dissipates away and $ \dot{\Phi} $ diminishes. This means
that when $ y^2 $ begins with a large value $ y^2 \sim 1 $ the dynamics
drives it towards smaller values.

\medskip

Even if initially $ y^2 >1 $
produces a non-inflationary stage [see eq.(\ref{ynfla})], this only occurs  for a
short period of time until $ y^2<1 $ where the evolution becomes inflationary.
The inflaton friction term continues to
dissipate away the kinetic energy and  when
$ y^2 = \mathcal{O}(1/N) \ll 1 $ the dynamics enters the slow roll inflationary
regime in earnest.

\medskip

We have restricted the above discussion to the case of  homogeneous
inflaton fields, where the energy is carried by the zero
mode of the inflaton up to small quantum fluctuations. 
However, a fast roll stage prior to slow roll
has also been studied in ref.\cite{tsu}, where a large amplitude
inhomogeneous condensate (tsunami inflation) was considered. In that
case modes with wavevectors of the order of the inflaton mass were
initially excited with large amplitude, the resulting
non-perturbative evolution of this initial state also leads to  a
fast roll stage which smoothly merges with the standard de Sitter
regime\cite{tsu}. The rapid redshift of non-homogeneous modes leads
to the formation of an effective homogeneous condensate after a few
e-folds. Therefore, a fast roll regime prior to the standard slow
roll regime  is a rather generic feature, either a result of an
almost equipartition between kinetic and potential energies for a
\emph{homogeneous} inflaton condensate, or from an inhomogeneous
non-perturbative condensate.

\subsection{Curvature perturbations during the fast roll stage}

For curvature perturbations, from eq.(\ref{sceq}) 
\be\label{WC}
W_\mathcal{R}(\eta) \equiv \frac1{z} \; \frac{d^2 z}{d \eta^2} =
\mathcal{V}_\mathcal{R}(\eta) +
\frac{\nu_\mathcal{R}^2-\frac14}{\eta^2} 
\ee 
where $ \nu_\mathcal{R}
= \frac32 + 3 \; \epsilon_v - \eta_v $ [see eq.(\ref{defnu})] and $
z $ is defined by eq.(\ref{za}).

\medskip

In order to compute $ W_\mathcal{R}(\eta) $, it is more convenient
to pass to cosmic time, in terms of which, 
\be \frac{d^2 z}{d
\eta^2} = a^2~(\ddot z~ +~ H \; \dot z) \; . 
\ee 
From eqs.(\ref{za}) and (\ref{WC}) and using the inflation equations of motion
(\ref{y})-(\ref{hpu}), the exact potential $ W_\mathcal{R}(\eta) $
can be written as 
\be\label{W} 
W_\mathcal{R}(\eta)  = C^2(\eta) \;
H^2 \left[ 2 - 7 \, y^2 + 2 \, y^4-   \textrm{sign}(\dot{\Phi})\,
\frac{2 \,  \sqrt2\; |y| \; V'}{M_{Pl} \; H^2}  - \frac{
V''}{H^2}\right] \; , 
\ee 
With the notation defined by eqs.(\ref{slowroll}) and (\ref{y}) we find, 
\be\label{Wa}
W_\mathcal{R}(\eta) = C^2(\eta) \; H^2\left[ 2 - 7 \, y^2 + 2\, y^4
- (3-y^2)(4 \; \sqrt{\epsilon_v} \; |y| \; \textrm{sign}(\dot{\Phi})
+ \eta_v) \right]       \; .
\ee
In order to clearly exhibit the natural scale of the potential $
W_{\mathcal{R}}(\eta) $ it is convenient to use the variables \cite{1sN}
\be\label{vars}
V= N \, m^2\, M_{Pl}^2 \; w(\chi)\quad ,\quad \Phi=
\sqrt{N} \,M_{Pl} \, \chi \quad  , \quad H = m \, h \, \sqrt{N}  \quad  ,\quad
t = \frac{\sqrt{N}}{m}~\tau \; ,
\ee
where $ N\sim 55 $ is the  number of efolds during slow roll and
$ m $ (the inflaton mass) defines the scale of the Hubble parameter 
during the stage of slow roll inflation.

This rescaling builds in the natural scales and
results in that $ w(\chi)\sim 1, \;  h\sim 1 $ during the slow roll stage
of inflation. Furthermore, as shown in ref.\cite{1sN},
the hierarchy of slow roll parameters is actually a
hierarchy in powers of $ 1/N $, for example
\be\label {eeta}
\epsilon_v = \frac1{2 \, N} \left(\frac
{w'}{w}\right)^2~~~~,~~~~ \eta_v = \frac1{N} \; \frac {w''}{w} \; .
\ee
In terms of these variables we obtain for the exact potential,
\bea\label{Wb}
W_\mathcal{R}(\eta) &=&  C^2(\eta)
\; h^2 \; m^2 \; N [ 2 - 7 \; y^2 + 2 \; y^4 - 2 \; \sqrt{\frac2{N}}
\frac{w'}{h^2} \; |y| \; \textrm{sign}(\dot{\Phi}) - \frac{ w''}{h^2 \; N}]
\quad , \cr \cr
y^2&=&3\left(1 - \frac {w}{3 \; h^2}\right)=
\frac{{\dot \chi}^2}{2 \; h^2 \; N} >0 \; .
\eea
displaying that for $ y\sim \mathcal{O}(1) $ the last two terms in
$ W_\mathcal{R}(\eta) $ eq.(\ref{Wb}) are
of order  $ \mathcal{O}(1/\sqrt{N})\ll1 $ and $ \mathcal{O}(1/N)\ll1 $ and can be neglected.

The above expressions in terms of the variable $ y$ are exact and allow to
analyze, besides slow roll inflation,  other regimes for inflation different
from slow roll. Recall the expression for $ W(\eta) $ in terms of the slow roll
parameters as given by :
\be \label {Wexsr}
W_{\mathcal{R}}(\eta)= a^2 \;  h^2 m^2 N^2 \; \left[2 + 2 \; \epsilon_v -
3 \; \eta_H + 2 \; \epsilon_v^2 - 4 \; \epsilon _v  \; \eta_H +
\eta_H^2 + \psi_H^2\right] \; ,
\ee
where, $ \eta_H = \eta_v - \epsilon_v, \; \psi_H = \psi_v - 3 \; 
\epsilon_v \; \eta_v +
3 \; \epsilon_v^2, \; \psi_v = \frac{1}{N^2} \; \frac {w'w'''}{w^2} $.
This expression is exact and appropriate in the slow roll approximation,
but it is not convenient in regimes different from slow roll.

In the slow roll approximation, \be y^2=\epsilon_v =
\mathcal{O}(1/N) \ll 1 ~~;~~C(\eta) = -\frac1{\eta \,H\,
(1-\epsilon_v)} \; , \ee and we recover \be W^{sr}_\mathcal{R}(\eta)
= \frac2{\eta^2}\Big[1+\frac32 \; (3 \; \epsilon_v-\eta_v)\Big]
\Rightarrow \mathcal{V}^{sr}_\mathcal{R}(\eta) =0 \; . \ee As shown
in eq.(\ref{ynfla}) the  range of the variable $ y^2 $ for
inflationary evolution is $ 0<y^2<1 $, which in turn implies:
\be\label{wh} 3 > \frac{w}{h^2} > 2 \quad {\rm or} \quad
\sqrt{\frac{w}3}<h<\sqrt{\frac{w}2} \; . \ee The expression of the
potential  eq.(\ref{Wb})  in terms of the variable $ y^2 $ is very
instructive. General properties of $ W(\eta) $, such as the sign of
the potential, can be analyzed from this expression revealing
different regimes. In the fast roll stage $ (y^2, \; w'/h^2,  \;
w''/h^2) = {\cal O}(1) $ and the dominant part of $ W(\eta) $ is
given by the polynomial in $ y $, the terms in the derivatives  $ w'
$ and $ w'' $ are of order $ O(1/\sqrt{N}) $ and $ O(1/N) $
respectively, namely: \be\label {Wd} W_\mathcal{R}(\eta) = C^2 H^2
\left[2 - 7 \; y^2 + 2\; y^4 + {\cal O}\left(\frac1{\sqrt{N}}\right)
\right] \ee The roots of $ W_\mathcal{R}(\eta) $ are up to
corrections $ {\cal O}\left(\frac1{\sqrt{N}}\right) $
$$
 y^2_- = \frac{7 - \sqrt{33}}4  = 0.31386\ldots \quad , \quad
y^2_+ = \frac{7 + \sqrt{33}}4  = 3.13859\ldots  \; .
$$
The potential $\mathcal{V}_\mathcal{R}(\eta) $ is obtained by
subtracting the slow roll contribution from $W(\eta)$, namely
\be \label{VV}
{\mathcal V}_\mathcal{R}(\eta)  =
W_\mathcal{R}(\eta) - \frac{2 + 9 \; \epsilon_v - 3 \; \eta_v +
{\cal O}\left(\frac1{N^2}\right)}{\eta^2}\,, \ee in the fast roll
stage \be \mathcal{V}_\mathcal{R}(\eta) = C^2 H^2 \left[ 2 - 7 \;
y^2 + 2\; y^4 + {\cal O}\left(\frac1{\sqrt{N}}\right)
\right] - \frac{2 + 9 \;  \epsilon_v - 3 \; \eta_v + {\cal
O}\left(\frac1{N^2}\right)}{\eta^2}  \; .
\ee
Thus, the full range $ 0\leq y^2 \leq 3 $, the range  $ y^2 <1 $ for which
inflation occurs and the roots of $ W(\eta) $ allow to identify
three different regimes:
\begin{itemize}
\item{  $ 0 < y^2 < y^2_- $,  in this region the potential
$ \mathcal{V}_\mathcal{R}(\eta) $ is repulsive and small.
This regime includes slow roll inflation for $ y^2 = {\cal
O}\left(\frac1{N}\right) \ll 1 $.}
\item{ $ y^2_- < y^2 < 1 $, corresponds to a  \emph{fast roll}
inflationary regime in which $ W_\mathcal{R}(\eta) $ is attractive
and consequently $ \mathcal{V}_\mathcal{R}(\eta) $ is attractive. }
\item{ $1< y^2 \leq 3 < y^2_+$ describes a fast roll but
non-inflationary regime in which the potentials $ W_\mathcal{R}(\eta) $
and  $ \mathcal{V}_\mathcal{R}(\eta) $ are attractive.}
\end{itemize}

In summary, when the initial value of $ y^2 $ is $ \gtrsim 1 $
the dynamics drives it monotonically towards smaller values.
The inflaton friction term continues to dissipate away the kinetic energy and when
$ y^2 < y^2_- $, the potential $ {\mathcal V}_{\mathcal R}(\eta) $
becomes repulsive but small and finally
when $ y^2 \ll y^2_- $ the dynamics enters the slow roll inflationary
regime in earnest. \\
Unless the initial conditions on the inflaton
determine that $ y^2 < y^2_- $, there is
\emph{always} a period of \emph{fast roll} inflation during which
the potential $ \mathcal{V}(\eta) $ for both curvature and tensor
perturbations is {\bf attractive}. As we will see below, this
attractive fast roll potential $ \mathcal{V}(\eta) $ produces a {\bf suppression} of the
quadrupole contributions to the angular power spectrum.

\subsection{Tensor perturbations during the fast roll stage}

The mode functions for tensor perturbations (gravitons) obey
eq.(\ref{sceq}) with
$$
W_T(\eta)\equiv C''(\eta)/C(\eta)  \; .
$$
Again, it is convenient to pass to cosmic time in terms of which,
\be\label{WCG}
W_T(\eta) = a^2(t) \; H^2(t)\Big[ 2 + \frac{\dot
H}{H^2}\Big] = C^2(\eta) \; H^2 \; ( 2 - y^2 ) \; ,
\ee
where we used the equation of motion (\ref{hpu}).

In the slow roll limit $ y = \epsilon_v = {\cal O}\left(\frac1{N}\right) \ll 1 , \;
{\mathcal V}_T(\eta) = 0 $ and eq.(\ref{eqnpsr2}) becomes a Bessel equation,
$$
\left[\frac{d^2}{d \eta^2} + k^2 -\frac{\nu_T^2-\frac14}{\eta^2}
 \right]S(k;\eta) = 0 \; ,
$$
where
$$
\nu_T = \frac32 + \epsilon_v +  {\cal O}\left(\frac1{N^2}\right)
\quad , \quad  W^{sr}_T(\eta) \simeq\frac{2 + 3 \;
\epsilon_v}{\eta^2}  \quad {\rm and} \quad {\mathcal V}^{sr}_T(\eta)
=0 \; .
$$
Notice that $ \nu_T $ differs from the index $ \nu_\mathcal{R} $ of
the scalar fluctuations at order $ {\cal O}\left(\frac1{N}\right)
$ [see eq.(\ref{defnu})].

\medskip

During the fast roll stage previous to the slow roll regime, $ y > 0 $
is not small and introduces an {\bf attractive} potential $ {\mathcal V}_T(\eta) $,
$$
{\mathcal V}_T(\eta) =  W_T(\eta) - \frac{2 + 3 \; \epsilon_v}{\eta^2} < 0 \; .
$$

\subsection{Fast roll in new and chaotic inflation}

\begin{figure}[h]
\begin{center}
\includegraphics[height=2.5in,width=2.5in,keepaspectratio=true]{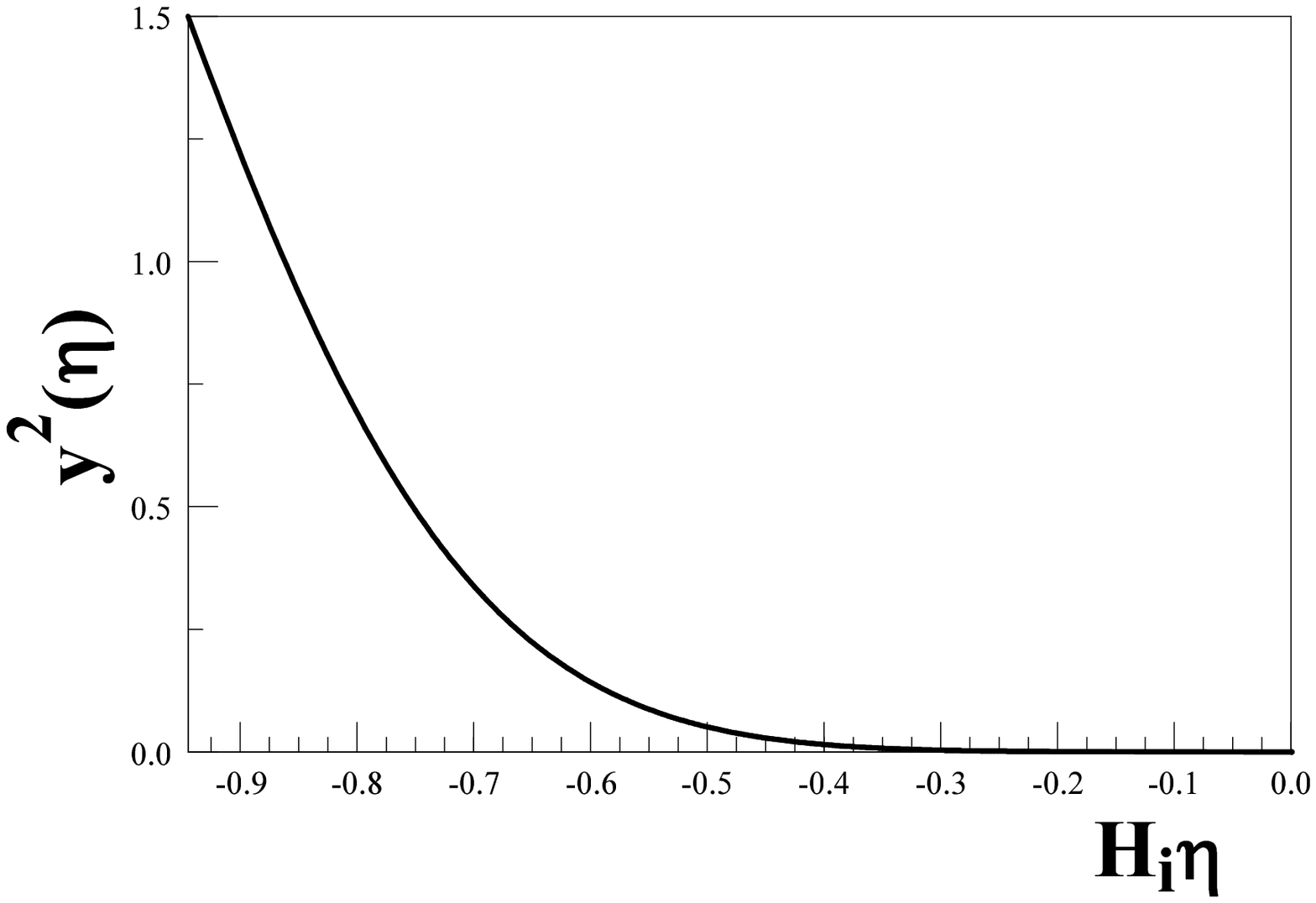}
\includegraphics[height=2.5in,width=2.5in,keepaspectratio=true]{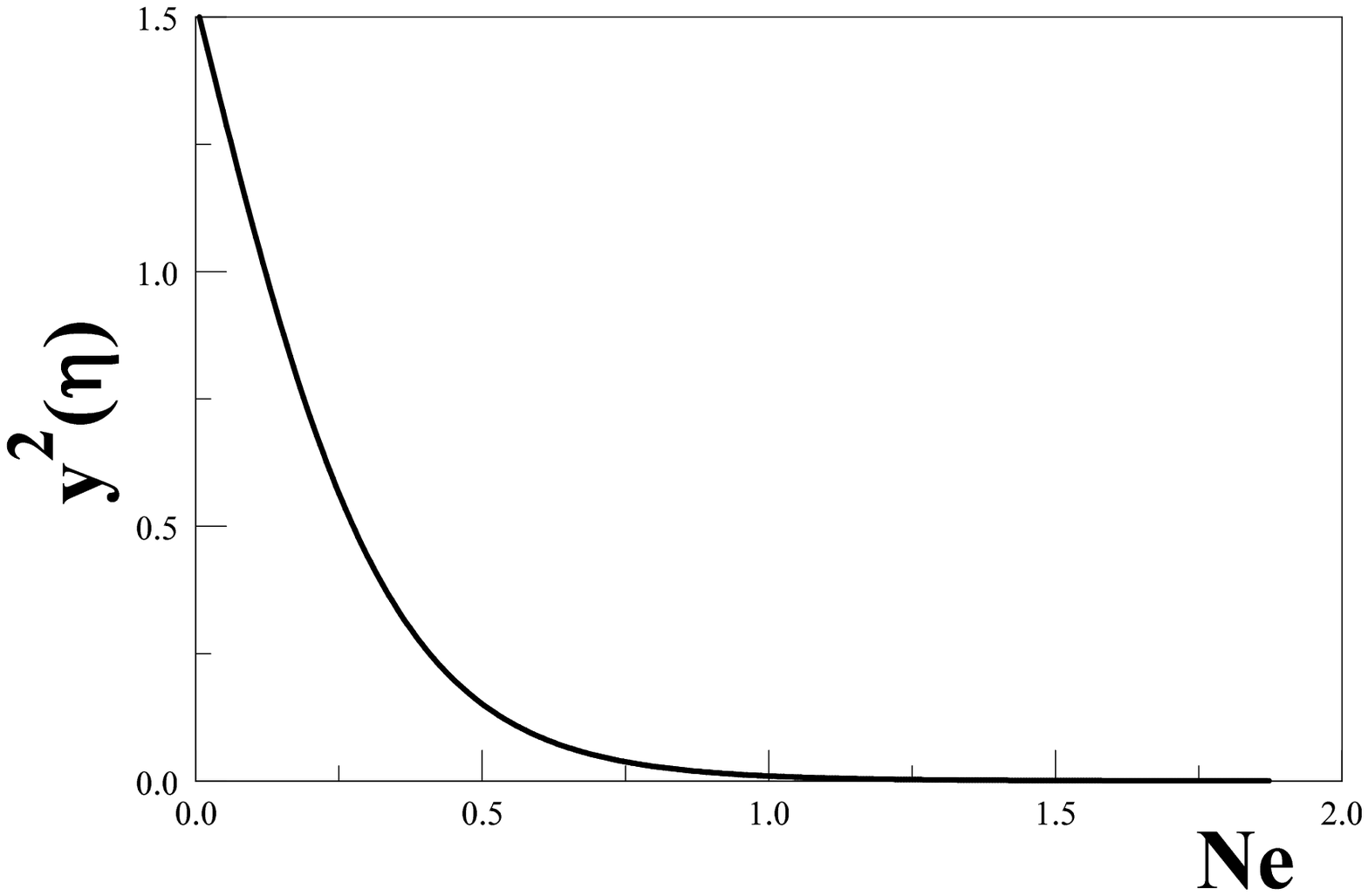}
\caption{ $ y^2(\eta)$ vs. $ \eta $ (left) and $ y^2(N_e) $ vs. $ N_e $
(right) for initial conditions with kinetic and potential inflaton
energy of the same order.  } \label{fig:Y}
\end{center}
\end{figure}

We consider models both of new inflation (small inflaton field) and
chaotic inflation (large inflaton field) to investigate the fast
roll dynamics prior to slow roll and its imprint on the quadrupole
mode as well as in the higher $ l$-modes. Let us focus first on new
inflation with the inflaton potential \be \label{simple} V(\Phi)=
V(0)\Big[1 - \lambda \; \frac{M^2_{Pl}}{m^2} \;
\frac{\Phi^2}{M^2_{Pl}} \Big]^2 ~~; ~~ V(0)\equiv 3 \;  H^2_{i} \;
M^2_{Pl}  \; , \ee where $ H_i $ is the Hubble parameter during slow
roll inflation. We note that during slow roll $ \lambda \;
\frac{M^2_{Pl}}{m^2 } = -\eta_v/4 $ and take $ \lambda \;
\frac{M^2_{Pl}}{m^2} = 0.008 $ as an example for numerical study. We
solve the equations (\ref{infa}) with the initial conditions $
\Phi(0)/M_{Pl}= 0~;~\dot{\Phi}^2(0)/[2 \; V_0] = 1~;~a(0)=1 $. These
initial conditions entail an \emph{equipartition} between the
kinetic and potential energy of the inflaton field at the initial
time.  Fig. \ref{fig:Y} displays $ y^2(\eta) $ (left panel) and $
y^2(N_e) $ (right panel) with $ N_e $ the number of e-folds from the
beginning of the evolution at $ t=0 $.

\medskip

These conditions initially yield $ y^2>1 $ which produces non-inflationary
dynamics, but after a very short time (about one e-fold)
$ y^2 $ drops below one and so inflationary dynamics begins in the
\emph{fast roll} regime $ y = {\cal O}(1) $, and after about one half e-fold when $ y^2
\sim 0.02 $ slow roll inflation begins in earnest.

\begin{figure}
\includegraphics[height=2in,width=2in,keepaspectratio=true]{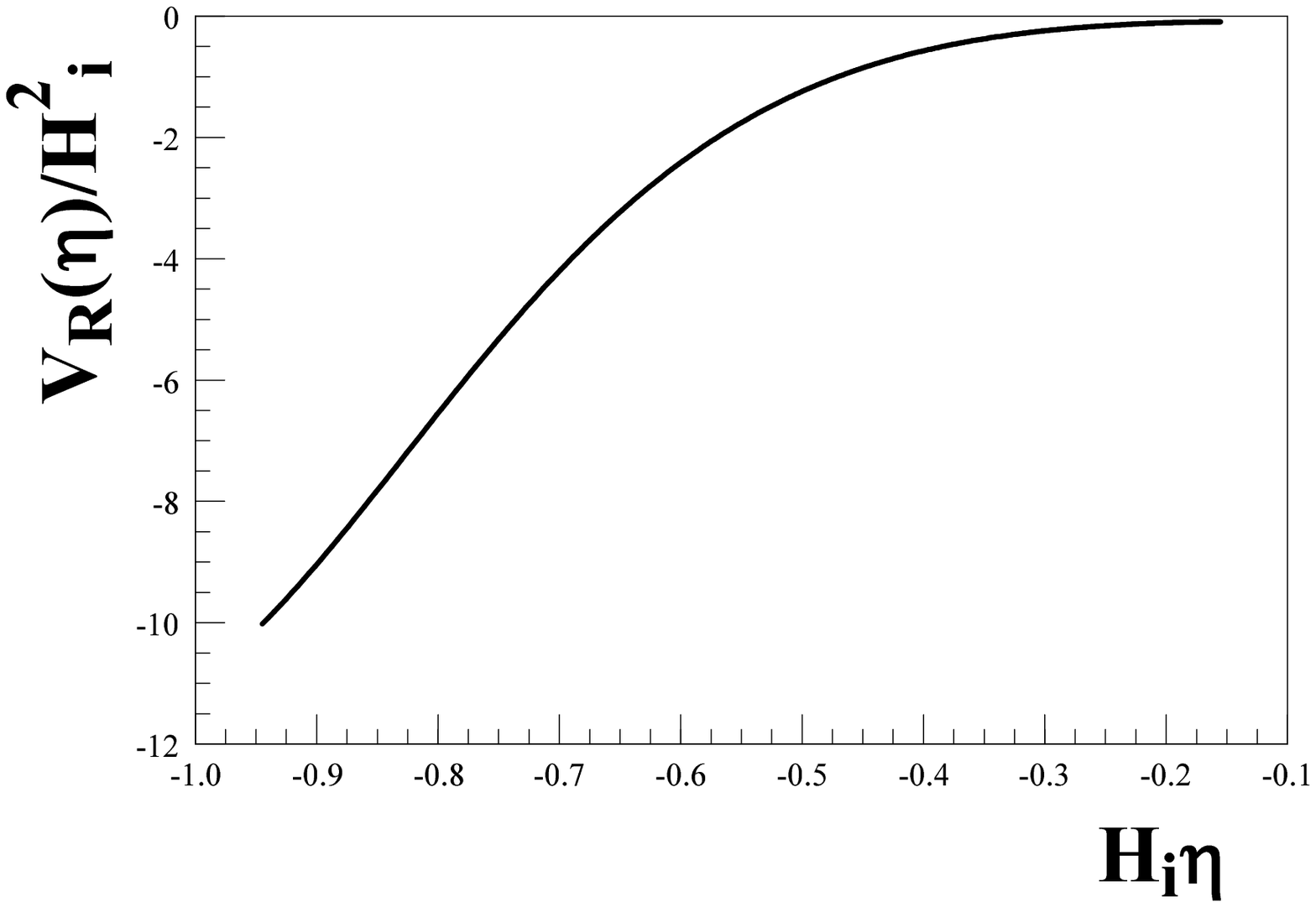}
\includegraphics[height=2in,width=2in,keepaspectratio=true]{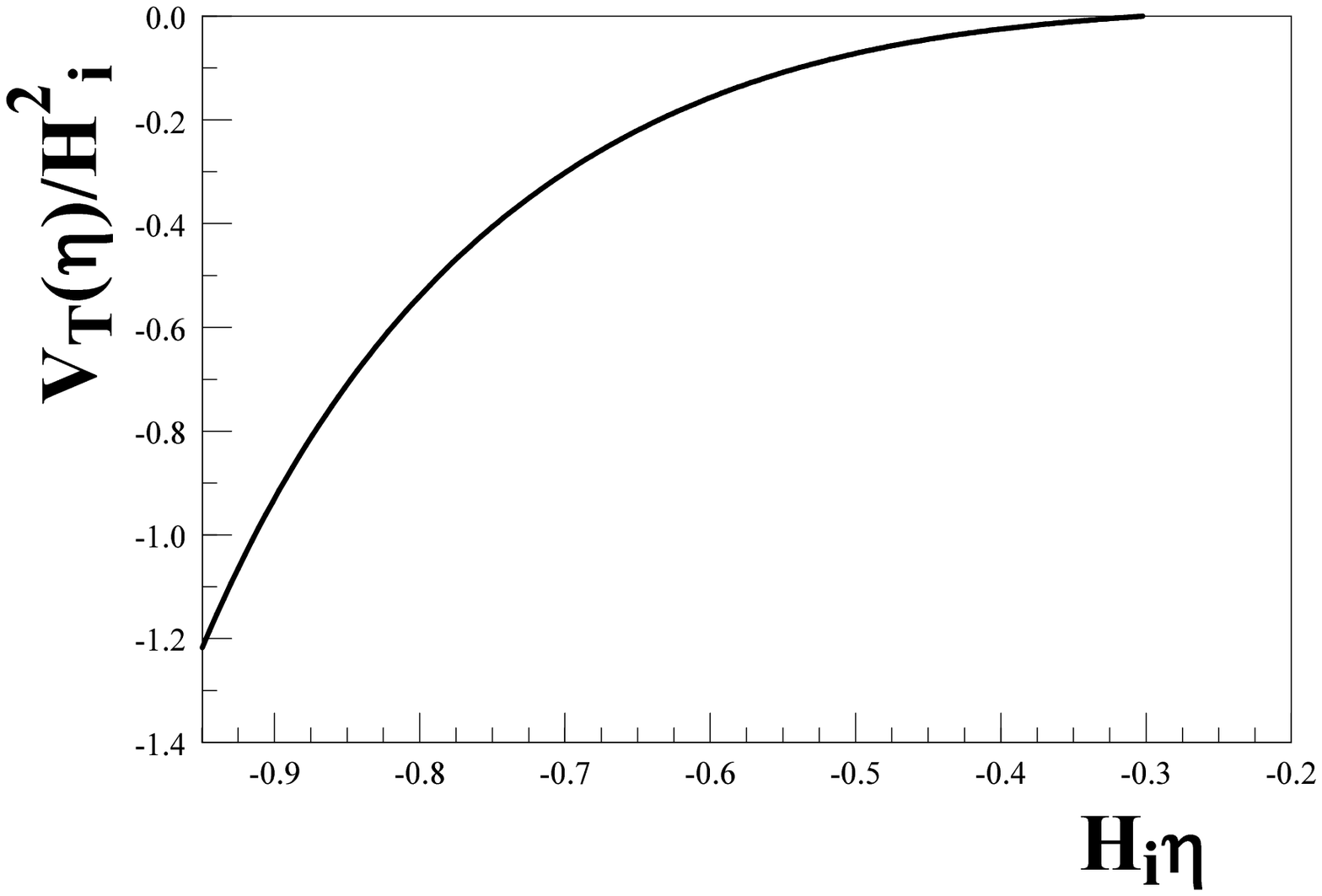}
\caption{ The potentials $ {\mathcal V}_\mathcal{R}(\eta)/H^2_{i} $
(left panel) and $ {\mathcal V}_T(\eta)/H^2_{i} $ (right panel) felt
by curvature and tensor perturbations respectively  vs $ H_i\,\eta
$, $ H_i $ being the Hubble parameter during the slow roll stage
(see fig.\ref{fig:hubble}). }\label{fig:potenciales}
\end{figure}

\begin{figure}
\begin{center}
\includegraphics[height=3in,width=3in,keepaspectratio=true]{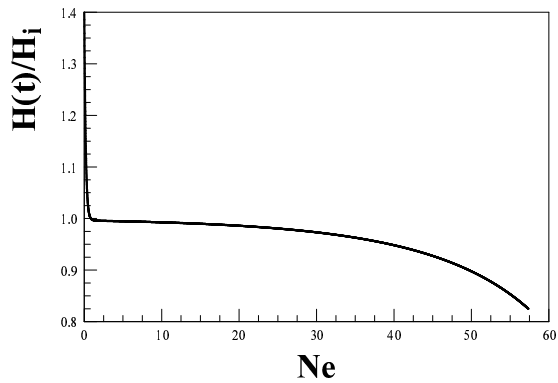}
\caption{$ H(t)/H_i $ vs number of e-folds} \label{fig:hubble}
\end{center}
\end{figure}

The potentials $ {\mathcal V}_\mathcal{R}(\eta) $ (left panel) and $
{\mathcal V}_T(\eta) $ (right panel) are shown in fig.
\ref{fig:potenciales}, and the evolution of the Hubble parameter is
displayed in fig. \ref{fig:hubble}. Figures (\ref{fig:Y}) and
(\ref{fig:potenciales}) show two distinct time scales: $ \eta_0
\approx -1/H_i $ at which the potential is localized and  features
its minimum, this is the beginning of the fast roll stage, and
$ \overline{\eta} \sim -0.3/H_i $ at which the potential vanishes,
$ y^2 \approx \epsilon_v $ and slow roll begins. The  brief \emph{fast
roll} stage is clearly seen from these figures to correspond to the
first e-fold of evolution. Fig. (\ref{fig:hubble}) confirms that the
fast roll stage lasts approximately one e-fold and that
$ \overline{\eta} $ corresponds to about 56-57 e-folds before the end
of inflation, namely $1-2$ e-folds before the modes corresponding to
today's Hubble radius exit the horizon during inflation.

For these parameters,   the height of the potentials are
approximately $ |\mathcal{V}_\mathcal{R}| \sim 10 \;
H^2_{i}~;~|\mathcal{V}_T| \sim 1.2 \; H^2_{i} $. The widths of the
potentials are approximately the same in both cases $ |\Delta /{
\eta_0}| \sim \Delta \,H_i \sim 0.2 $, see fig.
\ref{fig:potenciales}.

\begin{figure}
\includegraphics[height=2in,width=2in,keepaspectratio=true]{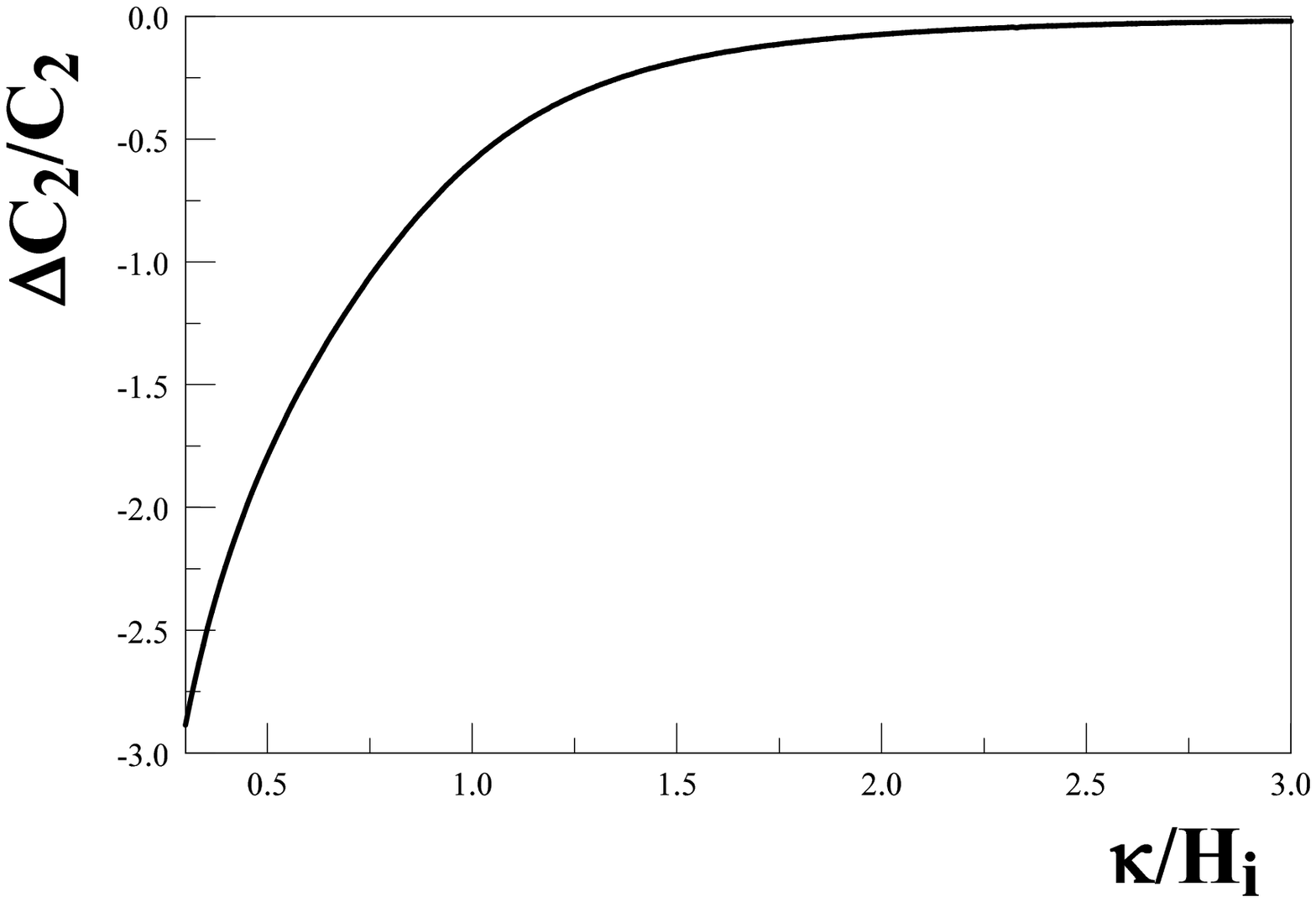}
\includegraphics[height=2in,width=2in,keepaspectratio=true]{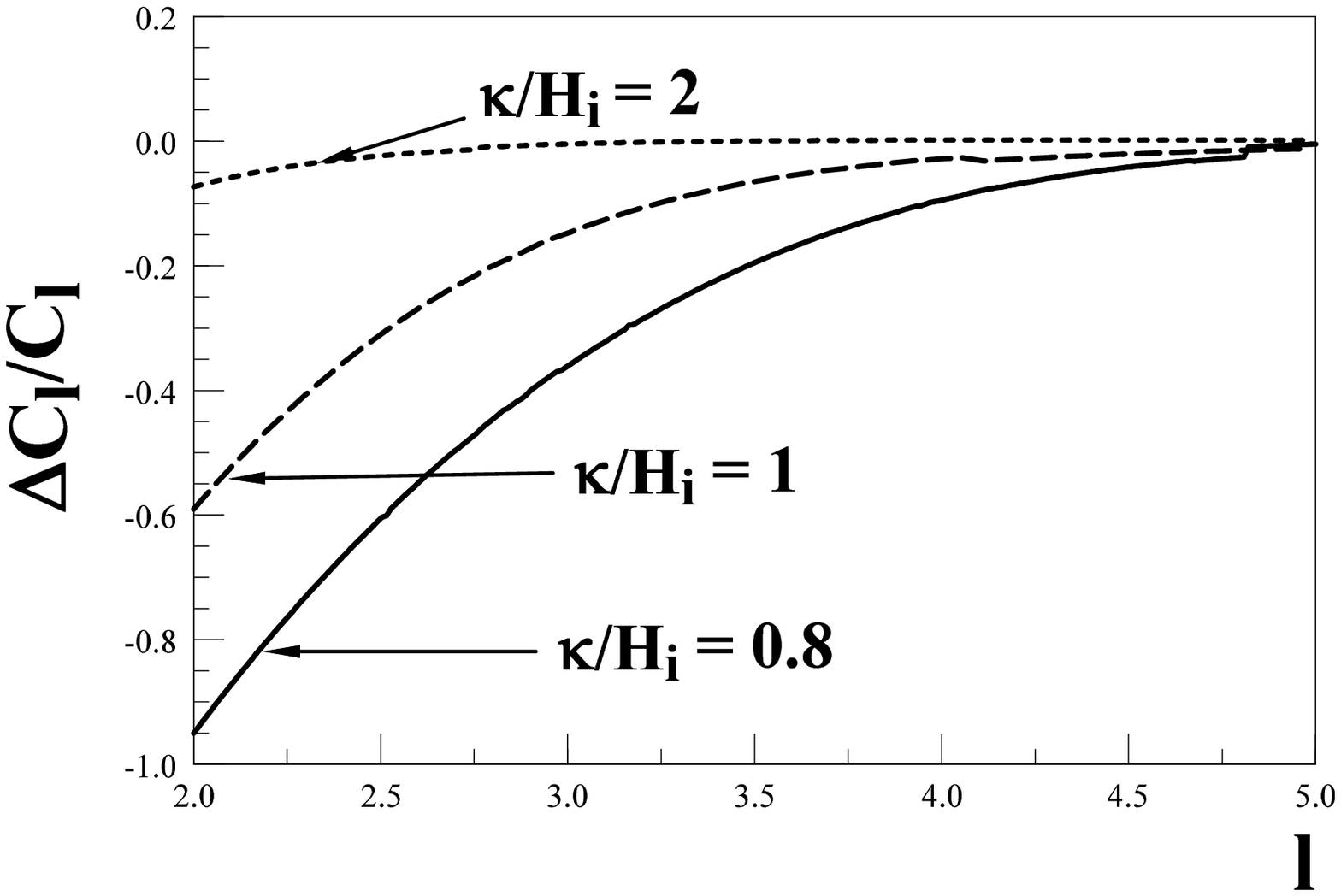}
\caption{ $ \Delta C_2/C_2 $ vs. $ \kappa/H_i $ (left panel) and $
\Delta C_l/C_l $ vs. $ l $ for $ \kappa/H_i=0.8, \; 1, \; 2 $
respectively for curvature perturbations.  $ \kappa = a_0 \, H_0/3.3
$}\label{fig:DCsnewinfla}
\end{figure}

\medskip

We have carried out analogous numerical studies in scenarios of
chaotic inflation with  similar results: if the initial kinetic
energy of the inflaton is of the same order as the potential energy,
a \emph{fast roll}   stage is \emph{always} present. The evolution
of $y^2$ and the potentials for curvature and tensor perturbations,
$ {\mathcal V}_\mathcal{R}(\eta)$ and  $ {\mathcal V}_T(\eta)$ are
again similar to those for new inflation and they are always
\emph{attractive} during the fast roll stage.

\medskip

An initial state for the \emph{inflaton} (inflaton classical
dynamics) with approximate \emph{equipartition} between kinetic and
potential energies is   a more \emph{general} initialization of
cosmological dynamics in the effective field theory than slow roll
which requires that the inflaton kinetic energy is much more smaller
than its potential energy. Therefore, we  conclude that the most
\emph{generic} initialization of the inflaton dynamics in the
effective field theory leads to a \emph{fast roll} stage followed by
slow roll inflation.

\section{Quadrupole Suppression}

In the Born approximation, the Bogoliubov coefficients eqs.(\ref{bofk}) 
are given by \cite{I},
\be
A(k)   =    1+  i \int^{0}_{-\infty}
   \mathcal{V}(\eta)\,|g_\nu(k;\eta)|^2 \, d\eta   \quad  , \quad
 B(k)  =  - i  \int^{0}_{-\infty}
   \mathcal{V}(\eta)\, g^2_\nu(k;\eta)  \, d\eta \; .\label{bofk0}
\ee
The transfer function of initial conditions given by eq.(\ref{DofkR}) can be computed
in the Born approximation, which is appropriate in this situation, using
eqs.(\ref{bofk0}) for the Bogoliubov coefficients $ A(k) $ and $ B(k) $,
\be 
D(k) = \frac1{k} \int^0_{-\infty} d\eta \;
\mathcal{V}(\eta) \left[\sin(2\, k \; \eta) \left(1  - \frac1{k^2 \,
\eta^2} \right)+ \frac2{k \, \eta} \, \cos(2\, k \, \eta) \right] \,
.\label{Dborn2} 
\ee 
The fractional change in the $ C_l's $ is
obtained by inserting this transfer function in the expression
(\ref{DelC}). We take the lower limit in the integral in
eq.(\ref{Dborn2}) to be $ \eta_0 \sim -1/H_i $ at which the fast
roll stage begins. The results of the numerical integrations for the
quadrupole $ l=2 $ and the higher multipoles are shown in
fig.\ref{fig:DCsnewinfla}.

The results displayed in this figure are strikingly similar to those
found in the examples studied in sections V.B and V.C of
ref.\cite{I} lending support to the conclusion that the  quadrupole
suppression as a consequence of  the attractive fast roll potential
$\mathcal {V} (\eta)$ is robust.

From eq.(\ref{kH}), the relevant dimensionless ratio $
\frac{\kappa}{H_i} $ that governs the multipole suppression $ \Delta
C_l / C_l $, is \be \frac{\kappa}{H_i} = \frac{a_{sr}}{3.3}
\label{rata} \; , \ee where $ a_{sr} $ is the scale factor when the
mode corresponding to the quadrupole wave vector $ k_Q $ exits the
Hubble radius during inflation.

We have fixed the initial value for the evolution to be at $
\eta=\eta_0 $ with $ C(\eta_0)\equiv 1 $, thus $ a_{sr} >1 $ is the
logarithm of the number of e-folds between the initial time of the
evolution and horizon crossing of $k_Q$. The left panel of fig.
\ref{fig:DCsnewinfla} clearly shows that the largest suppression for
the quadrupole corresponds to smallest values of $ a_{sr} $, with a
$ 10-20\% $ suppression for $ 1.7 \leq \kappa/H_i < 2.2 $. This
precisely corresponds to $ 3-4 $ e-folds between the onset of the
fast roll stage at $\eta_0$ and horizon crossing of the mode
corresponding to today's Hubble radius. The fast roll stage itself
only lasts about one e-fold and is followed by  slow roll.

\medskip

Thus, we conclude that there is a substantial suppression of the
quadrupole $ \sim 10-20 \% $  consistent with the observations, if $
k_Q $ exits the horizon within a 3-4 of e-folds after the
beginning of the slow roll stage, preceded by a short   fast roll
stage. Therefore, the observed quadrupole suppression is
successfully explained by the inflationary   dynamics --fast roll
followed by slow roll -- if inflation lasts not much more than
approximately $ N_{tot} \sim 59 $ e-folds.

\medskip

The similar form of the tensor potential $ \mathcal{V}_T $ leads to
a similar behavior in the change of the $ C_l's $ for the B-modes,
and the fractional change for the quadrupole of tensor modes is
smaller by almost an order of magnitude as gleaned from the
potentials displayed in fig. \ref{fig:potenciales}. This is a \emph
{general} prediction, again a consequence of a fast roll stage prior
to slow roll.

\medskip

A numerical analysis reveals that $ \Delta C_l / C_l \sim 1/ l^2 $
in agreement with the result of eq.(\ref{Cl10}), therefore the
suppression in the higher multipoles falls below the band of
irreducible cosmic variance and it is too small to be observable
within the present data.

A numerical fit to the curvature potential $
\mathcal{V}_\mathcal{R}(\eta) $  yields \be
\mathcal{V}_\mathcal{R}(\eta) \simeq \mathcal{V}_\mathcal{R}(\eta_0)
\, e^{-(\eta-\eta_0)/\Delta} \; ,  \label{expfit} \ee with $ \eta_0
\sim -1/H_i $ and $ |\Delta/\eta_0| \sim 0.2 $. With this analytic
expression which provides an excellent fit, we obtain   the
following asymptotic behavior of the transfer function   $
D_\mathcal{R}(k) $ and   distribution function $ N_\mathcal{R}(k) $
for large momenta: 
\be 
D_\mathcal{R}(k) \buildrel{k \to
\infty}\over= \frac{\mathcal{V}_\mathcal{R}(\eta_0)}{ 4\,k^2} \quad
, \quad N_\mathcal{R}(k) \buildrel{k \to \infty}\over=
\frac{\mathcal{V}^2_\mathcal{R}(\eta_0)}{16\,k^4} \label{nofknew} \; , 
\ee 
clearly indicating that these initial conditions are indeed
ultraviolet allowed and consistent with the form eq.(\ref{ocupa}).
We notice from figs. \ref{fig:Y} and \ref{fig:potenciales} that
indeed $ \mathcal{V}_\mathcal{R}(\eta) $ vanishes  when the slow
roll regime $ y^2 \ll 1 $ is reached.

\medskip

From eq.(\ref{VV}) with $ y^2(\eta_0) \sim 1 , \; C(\eta_0) = 1 $
and taking the initial conditions on the inflaton with approximate
equipartition between potential and kinetic energies, implies that $
H^2(\eta_0) \sim 2 \; H^2_{i} $ yielding \be
\mathcal{V}_\mathcal{R}(\eta_0) \sim - 10 \;  H^2_{i} \quad , \quad
\mathcal{V}_T(\eta_0) \sim - 2 \; H^2_{i} \; , \label{V0} \ee which
is consistent with  fig. \ref{fig:DCsnewinfla}. Comparing  with the
form eq.(\ref{ocupa}), and taking as an example $ N_\mu \sim 0.01 $,
indicates that the characteristic asymptotic k-scale $ \mu $ at
which the asymptotic form of  eq.(\ref{ocupa}) sets is $ \mu \approx 10~
H_i $, namely a few times the Hubble scale during slow roll
inflation. This shows that the energy scales involved in the
quadrupole suppression are of the \emph {same}  order as the scale
of inflation.

\vspace{1mm}

Therefore, the condition for observable suppression of the
quadrupole is that the modes with physical wavelengths of the order
of the Hubble radius today must cross the horizon during inflation
just $ 1-2 $ e-folds after the  beginning of the slow roll stage.
This condition  is easily understood from the approximate form
eq.(\ref{nofknew}) of the transfer function $ D(k)$. Since $ D(k)$
is strongly suppressed for $ k^2 \gg |\mathcal{V}| $, the potential
$ \mathcal{V}(\eta) $ will substantially influence the modes with
wavevector $ k $ if $ k^2 \lesssim |\mathcal{V}(\eta_0)| \sim 10 \;
H^2_{i} $. Since $ k = a_{sr} \;  H_i $, then clearly only $ 1-2 $
e-folds of evolution between the end of fast roll and the horizon
crossing lead to substantial effects on the mode functions from the
$ \mathcal{V}(\eta) $ potential.

\medskip

We have also studied chaotic inflationary scenarios with initial
conditions on the inflaton characterized by $ y^2 \sim 1 $ namely
with inflaton kinetic energy of the same order as the inflaton
potential energy. We find similar  results on the fractional
variation of low multipoles, the duration of the fast roll stage and
the scale of the fast roll potentials $
\mathcal{V}_\mathcal{R}(\eta)~,~\mathcal{V}_T(\eta)$   as for new
inflation.

\medskip

Therefore, we conclude that the phenomena associated with the fast
roll stage as a precursor to slow roll are robust, they \emph{do
not} depend on the inflationary model, but solely on the scale of
inflation and on approximate  equipartition between the kinetic and
potential energies in the initial condition for the classical
dynamics  of the inflaton. This  initialization of the inflaton
dynamics   and inflationary potentials that fulfill the slow roll
conditions generally guarantee that the dynamical evolution of the
inflaton features an initial fast roll stage that merges with the
usual slow roll inflationary stage. In turn, the fast roll stage
results in an attractive potential in the wave equations for the
mode functions of curvature an tensor perturbations, and a
consequent suppression of the quadrupole moment in their power
spectra.

\section{The Evolution of Perturbations as a Scattering Problem.}

The equivalence between the equations for the mode functions and the
Schr\"odinger equation with a potential allows us to bring to bear
the powerful results of potential scattering theory to
provide general statements on the properties of the solutions.

Eq.(\ref{sceq}) has the form of
the radial Schr\"odinger equation in the radial variable $ r\equiv -
\eta,  \; 0 \leq r  < \infty $ for the $L$-wave,  being $L$ a real
number, $ L \equiv \nu - \frac12 $ . We recognize in
eq.(\ref{eqnpsr2}) the centrifugal barrier
$$
\frac{\nu^2-\frac14}{\eta^2} = \frac{L(L+1)}{\eta^2}~~ ,~~
L \equiv \nu - \frac12 ,~~r\equiv - \eta.
$$
Thus, in the slow roll regime:
$$
L \equiv \nu - \frac12  = 1 +{\cal O}\left( \frac1{N}\right) \; .
$$
Eq.(\ref{eqnpsr2}) takes then the form
\be\label{rad}
\left[\frac{d^2}{dr^2}+k^2-\frac{L(L+1)}{r^2}- {\mathcal V}(r)
\right]f_{\nu}(k,r) = 0 \; .
\ee
The scattering solution of
eq.(\ref{rad}) with unit outgoing amplitude is defined by
\be
f_{\nu}(k,r)\buildrel{r \to +\infty}\over= e^{i k \, r}
\ee
This solution $ f_{\nu}(k,r) $ is called the Jost solution in scattering theory \cite{nt}, it is  identical to the Bunch-Davies initial conditions
eq.(\ref{fnuasy}) up to  a normalization factor $ \sqrt{2 \, k} $.

When $ {\mathcal V}(r) = 0 $ the Jost solution is given by
$$
f_{\nu}(k,r)_{\mathcal{V}=0} = i^{\nu+\frac12} \; \sqrt{\pi \; k \;
r} \; H^{(1)}_\nu(k\;r) \; .
$$
This function coincides with eq.(\ref{gnu}) up to a normalization
factor $ \sqrt{2 \; k} $. In particular,
\be \label{lib}
f_{\nu}(k,r)_{\mathcal{V}=0}\buildrel{r \to 0}\over=
\frac{\Gamma(\nu)}{\sqrt{\pi}} \; \left(\frac{k \; r}{2 \;
i}\right)^{\frac12-\nu} \; .
\ee
For $ r \to 0 $, eq.(\ref{eqnpsr2})
has two linearly independent solutions of the form: $ r^{\frac12 -
\nu} $ and $ r^{\frac12 + \nu} $; since $ \nu > 0 $ the first solution
dominates the behaviour of $ f_{\nu}(k,r) $ for $ r \to 0 $.

The Jost function of scattering theory is defined as the ratio
\be
F_{\nu}(k) \equiv \lim_{r \to 0} \frac{f_{\nu}(k,r)}{f_{\nu}(k,r)_{\mathcal{V}=0}} =
\frac{\sqrt{\pi}}{\Gamma(\nu)} \; \lim_{r \to 0} \left(\frac{k \,
r}{2 \, i} \right)^{\nu - \frac12} \; f_{\nu}(k,r) \; .
\ee

\subsection{ Scattering solutions and the primordial power}
By construction, the solution $ S(k;\eta) $ fulfils the Bunch-Davies
asymptotic condition
\be
S(k;\eta) \buildrel{\eta \to
-\infty}\over=\frac{e^{-ik\eta}}{\sqrt{2k}} \label{BDS}
\ee
This solution $ S(k;\eta) $ is proportional to the scattering Jost
solution as \be \label{reta} S(k;\eta) = \frac1{\sqrt{2k}} \;
f_{\nu}(k,r) \quad {\rm with}  \quad r = -\eta > 0 \quad . \ee It
can be shown on general grounds that $ f_{\nu}(k,r) $ is an analytic
function of $ k $ for Im$k > 0 $ and $ k \neq 0 $  \cite{nt}.
Moreover, $ k^{\nu - \frac12} \; f_{\nu}(k,r) $ as well as $ k^{\nu}
\; S(k;\eta) $ are analytic in a neighbourhood of  and including $ k
= 0 $.

For $ \eta \to 0^{-} $, eq.(\ref{eqnpsr2}) admits two independent solutions:
$ (-\eta)^{\frac12 - \nu} $ and $ (-\eta)^{\frac12 + \nu} $.
Since $ \nu > 0 $, the first solution is the irregular one for
$ \eta \to 0^{-} $ and it dominates over the regular solution
$ (-\eta)^{\frac12 + \nu} $. \\
The $ \eta \to 0^- $ behaviour of the modes in the $
\mathcal{V}(\eta) \equiv 0 $ case is given by eq.(\ref{geta0}),
while in the general case $ \mathcal{V}(\eta) \neq 0 $ we have
\be\label{sf}
 S(k;\eta)\buildrel{\eta \to 0^-}\over= \frac{\Gamma (\nu)}{\sqrt{2 \,\pi \; k}}
\; \left(\frac2{i \; k \; \eta} \right)^{\nu - \frac12} \;
F_{\nu}(k) \; ,
\ee
where $ F_{\nu}(k) $ stands for the Jost
function. It follows that $ F_{\nu}(k) $ is analytic  for Im$k > 0 $
and \cite{nt}
\be\label{Flib}
\lim_{k \to \infty}  F_{\nu}(k) = 1 \; .
\ee
The primordial power spectra are given by eqs.(\ref{curvapot}).
Eq.(\ref{potBD}) for Bunch-Davies (BD) initial conditions is
valid when $ \mathcal{V}(\eta) = 0 $ and the mode functions
  behave as in eq.(\ref{geta0}) for $ \eta \to 0^- $. From
eqs.(\ref{geta0}) and (\ref{sf}) we find for $ \mathcal{V}\neq 0 $,
\be \frac{|S(k;\eta)|^2}{ |g_\nu(k;\eta)|^2}\buildrel{\eta \to
0^-}\over=  |F_{\nu}(k)|^2 \quad . \ee Therefore, we find the
equivalence, \be\label{potjo} \frac{P(k)_{\mathcal{V}}}{ P^{sr}(k)}=
|F_{\nu}(k)|^2 = 1+ D(k) \quad . \ee Namely,    $ |F_{\nu}(k)|^2 $
yields the change in the primordial power spectrum due to the
potential $ \mathcal{V}(\eta) $. This is an  important result, which
allows to obtain general information on the transfer  function of
initial conditions $ D(k) $ from established results of potential
scattering.

We obtain the Jost function  $ F_{\nu}(k) $ from the $ \eta \to 0^-
$ behavior of eq.(\ref{solu}) with the result
\be\label{FJO}
 F_{\nu}(k) = 1 + i^{\frac12-\nu} \; \sqrt{\pi} \int^0_{-\infty}
\sqrt{-\eta} \; d\eta \; J_{\nu}(-k \, \eta) \; \mathcal{V}(\eta) \;
S(k;\eta) \quad .
\ee
where $ J_{\nu}(z) $  is Bessel's function.

In the scale invariant case $ \nu = \frac32 $ the Jost function
takes the simpler form
\be \label{f32}
 F_{\frac32}(k) = 1 - i \; \sqrt{\frac2{k}}\int^0_{-\infty} d\eta \;
\left[\frac{\sin(k \, \eta)}{k \, \eta} - \cos(k \, \eta) \right] \;
\mathcal{V}(\eta) \; S(k;\eta) \quad .
\ee
The large $ k $ behavior
of the Jost solutions and Jost functions follows by solving
eqs.(\ref{solu}) and (\ref{FJO}) by iteration. To dominant order we
find that the Jost solution is given by the $ {\mathcal V(\eta)}=0 $
solution eq.(\ref{gnu}) while the Jost function equals unity [see
eq.(\ref{Flib})].

The logarithm of the Jost function has the following asymptotic expansion
around $ k = \infty $ \cite{ist},
$$
 \log F_{\nu}(k) = - \sum_{n=1}^{\infty} \frac{c_n}{(2\, i \; k)^n} \; ,
$$
where the $ c_n $ are {\bf real} coefficients functionals of the
potential $ \mathcal{V}(\eta) $. The first  coefficients take the
form,
$$
c_1 = \int^0_{-\infty} d\eta \; \mathcal{V}(\eta) \quad ,  \quad c_2
= \mathcal{V}({\bar \eta}) \quad .
$$
Therefore, 
\be 
\log |F_{\nu}(k)|^2 = \frac{\mathcal{V}({\bar \eta})}{2 \;
k^2} + {\cal O}\left(\frac1{k^4} \right) \quad . 
\ee 
We see that
asymptotically $ |F_{\nu}(k)|^2 < 1 $ for a potential which is
attractive at the end of fast roll [$ \mathcal{V}({\bar \eta}) < 0 $]. Combined
with eq.(\ref{potjo}) this result shows in general that  an
attractive potential $ \mathcal{V}(\eta) $ {\bf suppresses} the
primordial power.

\bigskip

Computing the $ \eta \to 0^- $ behavior of $ S(k;\eta) $ from
eq.(\ref{solSR}) permits to relate  the Bogoliubov coefficients $
A(k) $ and $ B(k) $ with the Jost function as
\be
A(k)+ i^{1-2 \, \nu} B(k) = F_{\nu}(k) \quad .
\ee
where we used eqs.(\ref{gnu}) and (\ref{sf}).

Therefore,
\be \label{relFD}
|F_{\nu}(k)|^2 - 1 = 2 \;  |B(k)|^2
-  2 \; {\rm Re}~[A(k) \,  B^*(k) \; i^{2 \, \nu-3} ]  = D(k) \; .
\ee
and we recover the transfer function for the initial conditions
$ D(k) $ introduced in ref.\cite{I}. Using eq.(\ref{bogonum}), eq.(\ref{relFD}) reduces
\emph{exactly} to eqs.(\ref{DofkR}).

For large $ k $, the mode functions $ S(k;\eta) $ as well as the $
g_\nu(k;\eta) $ tend to their plane wave asymptotic behaviour
$$
 S(k;\eta)\buildrel{k \to \infty}\over= g_\nu(k;\eta)\buildrel{k \to \infty}\over=
 \frac{e^{-ik\eta}}{\sqrt{2k}} \; .
$$
A look at eq.(\ref{solSR}) shows that this implies $ B(\infty) = 0,
\; A(\infty) = 1 $. More precisely, we find from eq.(\ref{bofk}),
\be
A(k)  \buildrel{k \to \infty}\over=   1+ \frac{i}{2
\; k} \int^{0}_{-\infty} \mathcal{V}(\eta) \; d\eta \quad ,  \quad
 B(k)  \buildrel{k \to \infty}\over=  -\frac{i}{2 \; k} \int^{0}_{-\infty}
e^{-2 \, i \, k \; \eta} \; \mathcal{V}(\eta) \;  d\eta
\ee
According to the Riemann-Lebesgue lemma, $ B(k) $ vanishes for $ k
\to \infty $ faster than any negative power of $ k $. Hence, the
convergence at large $k$ in the integrals for the energy momentum
tensor is guaranteed.

\bigskip

The Bogoliubov coefficients $ A(k) $ and $ B(k) $ are related to the
usual transmission ($T$) and reflection ($R$) coefficients of
scattering theory by the relation,
\be
T(k) = \frac1{A(-k)}~~;~~ R(k) = \frac{B(-k)}{A(-k)} \quad \label{rela}
, \quad  |R(k)|^2  +  |T(k)|^2 = 1 \, .
\ee
We provide with Table I a dictionary to translate from the
fluctuations language to the scattering framework.

\bigskip

\begin{tabular}{|c|c|}\hline
  & \\ Fluctuations   & Scattering Problem \\
   & \\ \hline
 $ \qquad  -\infty < \eta < 0 \; $   & $ \qquad \; 0 < r < \infty  $ \\ \hline
Bunch-Davies initial conditions: & Jost solutions: \\    &  \\
 $ S(k;\eta) = \frac{e^{-i k \eta}}{\sqrt{2k}} \;    $ for $ \eta \to -\infty $
& $ f_{\nu}(k,r) = e^{i k \, r} $ for $ r \to \infty $ \\ \hline
Superhorizon modes:
& Jost Function: \\ & \\
   $ S(k;\eta) \buildrel{\eta \to  0^-}\over\sim (-\eta)^{\frac12 - \nu} $
& $ F_{\nu}(k) \equiv \frac{\sqrt{\pi}}{\Gamma(\nu)} \; \lim_{r \to
0} \left(\frac{k \, r}{2 \, i} \right)^{\nu - \frac12} \;
f_{\nu}(k,r) $
\\ \hline
$ \qquad $    Power spectra & Modulus Square of the Jost Function =\\
$\frac{ P_{\mathcal{V}}(k)}{ P^{sr}(k)}$ & $ =  | F_{\nu}(k) |^2 $
\\ \hline
\end{tabular}

\bigskip

{TABLE 1. Correspondence between the scalar fluctuations as
functions of the conformal time $ \eta < 0 $ and the radial wave
functions, of $ r > 0 $ and angular momentum $ L \equiv \nu -
\frac12 $.}

\subsection{The quadrupole suppression: General results}

\begin{figure}[h]
\begin{center}
\begin{turn}{-90}
\centering
\psfrag{psix}{$ \Psi(x) $ vs. $ x $}
\includegraphics[height=10cm,width=10cm,keepaspectratio=true]{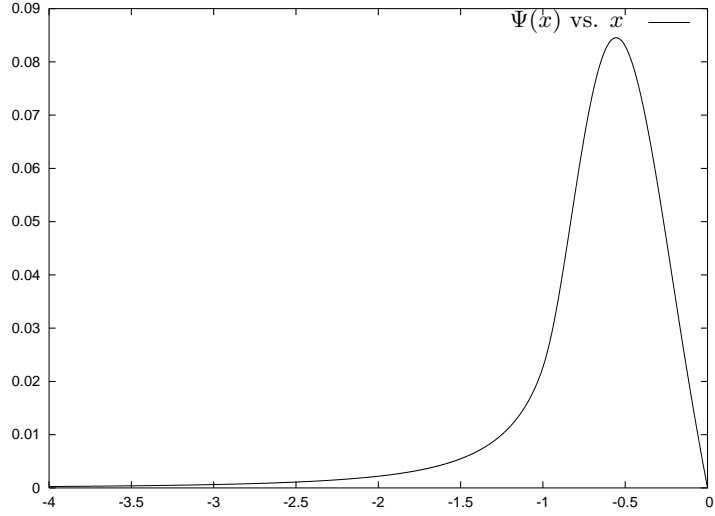}
\end{turn}
\caption{ The odd function $ \Psi(x) $ vs. $ x $ for negative $ x $ [see
eq.(\ref{defPsi})]. This function convoluted with the potential $
\mathcal{V}(\eta) $ yields the change on the quadrupole $
\frac{\Delta C_2}{C_2} $ [see eq.(\ref{cuaB})].} \label{Psi}
\end{center}
\end{figure}

We now implement the exact relations between the scattering
problem and the power spectra of perturbations derived in the previous subsection to obtain general
results for the quadrupole produced by the potential $
\mathcal{V}(\eta) $ . From eq.(\ref{DelC})
for $ l = 2 $ and to zeroth order in slow roll, the fractional change
in the quadrupole is given by,
\be \label{DC2}
\frac{\Delta C_2}{C_2} =
\frac{\int^\infty_0 D(\kappa\,x)~ f_2(x)\,dx}{\int^\infty_0
f_2(x)\,dx} = 3 \int^\infty_0 \frac{dx}{x} \; D(\kappa\,x)~ \left[
j_2(x) \right]^2 \; ,
\ee
where $ j_2(x) $ is the spherical Bessel function of order two \cite{abra}.
We  compute the
transfer function $ D(k) $ from the Jost function using
eqs.(\ref{f32}) and (\ref{relFD})in the Born approximation, which turns be an excellent one for this purpose;
since in fact the potential $ \mathcal{V}(\eta) $ is small. The Jost function in the
Born approximation to zeroth order in slow roll is given by
$$
F_{\frac32}(k) = 1 + \frac{i}{2 \; k}\int^0_{-\infty} d\eta \,
\mathcal{V}(\eta) \left(1 - \frac{i}{k \, \eta}\right) \left[1 +
e^{-2 \, i \, k \, \eta}-\frac{1- e^{-2 \, i \, k \, \eta}}{i \, k
\, \eta} \right] \, .
$$
Therefore up to first order in $ \mathcal{V}(\eta) $ (Born
approximation) we find
$$
D(k) = |F_{\frac32}(k)|^2 - 1 = \frac1{k} \int^0_{-\infty} d\eta
\;  \mathcal{V}(\eta) \left[\sin(2\, k \; \eta) \left(1  -
\frac1{k^2 \, \eta^2} \right)+ \frac2{k \, \eta} \, \cos(2\, k \,
\eta) \right] \, .
$$
Inserting this expression for $ D(k) $ into eq.(\ref{DC2}) yields
\be\label{cuaB}
\frac{\Delta C_2}{C_2} = \frac1{\kappa}
\int^0_{-\infty} d\eta \;  \mathcal{V}(\eta) \; \Psi(\kappa \; \eta)
\ee where
\be \label{defPsi}
\Psi(x) \equiv 3 \int_0^{\infty} \frac{dy}{y^4} \left[ j_2(y) \right]^2
\left[ (y^2 -\frac1{x^2} )\sin(2\, y \, x )+ \frac{2 \,
y}{x} \; \cos(2\, y \, x ) \right]
\ee
$ \Psi(x) $  is an odd function of $ x $. The
integral in eq.(\ref{defPsi}) can be computed in terms of elementary
functions by using the power series expansion \cite{gr}
$$
\left[ j_2(x) \right]^2 = \frac{\sqrt{\pi}}2 \sum_{k=0}^{\infty}
\frac{(-1)^k \; x^{2\,k + 4}}{ k! \; \Gamma(k + \frac72 ) \;
(k+3)(k+4)(k+5)} \; ,
$$
with the result
\bea
&&\Psi(x)=- \frac3{4 \; x^3} \sum_{k=0}^{\infty} \frac1{x^{2 \, k}}
\frac1{(k + \frac32)(k + \frac52)(k+4)(k+5)} \left[1 +
\frac1{(k + \frac12)(k + 3)} \right] = \cr \cr
&& =- \frac1{x^3} \sum_{k=0}^{\infty} \frac1{x^{2 \, k}} \left[
\frac1{105} \frac1{k + \frac12}+\frac1{35} \frac1{k + \frac32}
- \frac15 \frac1{k + 3}+ \frac9{35} \frac1{k + 4} - \frac2{21}
\frac1{k + 5}   \right] \; .
\eea
These series can be summed up explicitly with the result
\be
\Psi(x)= \frac1{105 \; x^2} \left[ p(x) \;
(1-x)^3 \; \log\left|1 - \frac1{x} \right| -  p(-x) \; (1+x)^3 \;
\log\left|1 + \frac1{x} \right|\right] + \frac2{105 \, x} - \frac{13
\, x}{126} + \frac{22 \, x^3}{105} - \frac{2 \, x^5}{21}
\ee
where $ p(x) $ is the sixth order polynomial
$$
p(x)\equiv 10 \, x^6 + 30 \, x^5 + 33 \, x^4 + 19 \, x^3 + 9 \, x^2
+ 3 \, x + 1 \; .
$$
The function $ \Psi(x) $ is negative for $ x > 0 $ and positive for
$ x < 0 $. It vanishes for $ x \to 0 $ and for  $ x \to \infty $ as,
$$
\Psi(x)\buildrel{x \to  0}\over= - \frac{x}6 + {\cal O}(x^3) \; .
$$
and
$$
\Psi(x)\buildrel{x \to \infty}\over= - \frac1{60 \; x^3} +
{\cal O}\left(\frac1{x^5}\right) \; .
$$
Fig. \ref{Psi} displays   $ \Psi(x) $ as a function of $ x $ for
negative $ x $ . $ \Psi(x) $ features a  maximum at $ x = x_M =
-0.555\ldots $ with $ \Psi(x_M) = 0.08453\ldots $.

Eq.(\ref{cuaB})  highlights the {\bf general} result that an {\bf attractive} 
potential $ \mathcal{V}(\eta) < 0 $ yields a  {\bf suppression} of the
quadrupole since $ \Psi(x)>0 $ for negative values of its argument $ x $.

\medskip

These   results establish  unequivocally that the attractive
potentials $ \mathcal{V}_\mathcal{R}(\eta)~;~\mathcal{V}_{T}(\eta)
$, which are a consequence of the fast roll stage,  lead  to a
suppression of the quadrupole for curvature and tensor
perturbations.

\subsection{The Inverse Problem. Reconstructing the fast roll potential
$ \mathcal{V}(\eta) $ from the primordial power}

In scattering theory, the potential can be obtained from the
scattering data, through the Gelfand-Levitan equation. This is a
linear integral equation which determines the potential $ \mathcal{V}(r) $
from the knowledge of the modulus of the Jost function and the bound states\cite{ist}.

The Gelfand-Levitan equation can be written as
\be\label{GL}
K_{\nu}(r,r') + G_{\nu}(r,r') + \int_0^r dr'' \; K_{\nu}(r,r'') \;
G_{\nu}(r'',r') = 0 \; . \ee where $ G_{\nu}(r,r') $ is a known
function that can be expressed in terms of the Jost function as
follows \be\label{GL2} G_{\nu}(r,r') = \sqrt{r \; r'}\int_0^{\infty}
k \; dk \; J_{\nu}(k \; r)\; J_{\nu}(k \; r') \left[
\frac1{|F_{\nu}(k)|^2}-1\right] \; ,
\ee
where the $ J_{\nu}(z) $
are Bessel functions, and the kernel $ K_{\nu}(r,r') $ is obtained
by solving eq.(\ref{GL}). Once $ K_{\nu}(r,r') $ is computed, the
potential follows as
\be\label{GL3}
\mathcal{V}(r) =2 \; \frac{d}{dr} K_{\nu}(r,r) \quad .
\ee
Eq.(\ref{GL}) is the Gelfand-Levitan
equation in absence of bound states. By bound states we mean
solutions of eq.(\ref{rad}) which are regular at $ r = 0 $ and decay
exponentially for $ r \to \infty $. We will not consider their
presence since the analysis in secs. II and III of ref.\cite{I}
indicates that bound states are absent in the present case.

\medskip

We have seen in eq.(\ref{potjo}) that the deviation of the
primordial power from slow roll is given by the square modulus of
the Jost function. Eqs.(\ref{GL})-(\ref{GL3}) show that this
deviation from the BD-slow roll primordial power explicitly
determines the potential $ \mathcal{V}(\eta) $. The
present quantitative information about the deviation of the
primordial power from slow roll is too scarce to feed back into the
Gelfand-Levitan equation, but it is important to see that the fast roll potential
$ \mathcal{V}_{\mathcal R}(\eta) $ felt by the fluctuations and hence $ W_{\mathcal R}(\eta) $ can be
explicitly determined from the primordial power data.

\section{Conclusions}

Although the latest analysis of the WMAP data confirms the basic
paradigm of slow roll inflation and renders much less statistical
significance to potential departures from its basic predictions, the
  anomalously low quadrupole in the CMB remains a long-standing
challenge.

In this article we proposed a mechanism that yields a suppression of
the low multipoles both for curvature and tensor perturbations,
\emph{within the effective field theory of inflation}. The main
premise of our observation is that a more general initialization of
the classical dynamics of the inflaton, allowing for approximate
equipartition between initial kinetic and potential energy of the
inflaton leads to a brief period of \emph{fast roll dynamics} that
is the precursor to the usual slow roll stage. This early fast roll
stage results in an \emph{attractive potential} in the wave equation
for the mode functions of curvature and tensor perturbations.
Implementing the methods and borrowing   the results from our
companion article\cite{I}, we show that this attractive potential
yields a transfer function for initial conditions $D(k)$ which
fulfills the stringent criteria of renormalizability and small back
reaction and yields a $10-20\%$ suppression of the CMB  quadrupole
consistent with the observational data. We also predict a small
$\sim 2-4\%$ quadrupole suppression for B-modes.

Our main results are summarized as follows:

\begin{itemize}

\item{Within the framework of the effective field theory of
inflation at the GUT scale we show that allowing for an initial
state of the inflaton for which its kinetic energy is of the same
order as the potential energy, there emerges a brief stage prior to
slow roll in which the inflaton rolls \emph{fast}. We call this
brief, but consequential stage, the \emph{fast roll regime}. The
inflaton potential fulfills the slow roll conditions and is the {\bf
same} both in the slow roll and in the fast roll regime. We prove
that this brief   fast roll stage generates an \emph{attractive}
localized potential for the mode functions of metric and tensor
perturbations.  Such potential leads to initial conditions for the
fluctuations \emph{during the slow roll stage} which are different
from Bunch-Davies and are consistent with renormalization and
negligible backreaction.}

\item{We provide an exhaustive numerical analysis for several inflationary
models with the result that for generic inflaton initial conditions
with \emph{equipartition} between kinetic and potential inflaton
energy there is a brief period of {\bf fast roll} that lasts
approximately $ \sim 1 $ e-fold. This brief stage translates in a
potential $\mathcal{V} (\eta) $ in the wave equation for the mode functions of 
curvature and tensor perturbations. The typical scales of these potentials are
$ \mathcal{V}_\mathcal{R} \sim -10 \; H^2_{i} $ for curvature
perturbations and $ \mathcal{V}_T \sim -2 \; H^2_{i} $ for tensor
perturbations, where $H_i$ is the Hubble parameter \emph{during slow
roll inflation}. A suppression of the CMB quadrupole of about
$10-20\%$, consistent with observation is obtained if the mode
corresponding to the quadrupole, whose physical wavelength is of the
order of the Hubble radius today,  crossed the horizon within $ 2-3
$ efolds after the  beginning of slow roll stage.}

\item{Our study also predicts a suppression of the quadrupole for the B-modes,
with a fractional change of at least an order of magnitude smaller
than that for temperature fluctuations.}

\item{ The evolution of the inflationary perturbations has been shown to be 
equivalent to the scattering by a
potential and useful expressions between the two sets of solutions
and observables have been derived. By implementing the methods of
scattering theory we prove in general that the   CMB quadrupole is
suppressed by the attractive potential $\mathcal{V}(\eta)$ which is
a consequence of the fast roll stage.  }

\item{Thus, we conclude that generic ultraviolet-finite
initial conditions imprinted upon gaussian curvature perturbations
from a fast roll stage just prior to slow roll inflation
successfully explain the low quadrupole. Such suppression happens
provided the inflationary stage does not last more than $ \sim 57-58
$ e-folds. Therefore this suppression mechanism successfully
explains the low CMB quadrupole provided there is  the \emph {upper
bound} $ N_{tot} \sim  N_Q + 4 = 59 $ on the total number of efolds during
inflation. This upper bound results from the following accounting:
the modes corresponding to the quadrupole crossed out of the Hubble
radius during the slow roll stage approximately $ N_Q = 55$ e-folds before
the end of inflation. However for the potential
$ \mathcal{V}_\mathcal{R}(\eta) $ to influence these modes, the exit
time cannot be more than approximately $3$ e-folds after the end
of the fast roll stage, which itself lasts approximately $1$ e-fold,
yielding a total of about $ N_{tot} = N_Q + 4 = 59$ e-folds. }

\end{itemize}

\begin{acknowledgments} D.B.\ thanks the US NSF for support under
grant PHY-0242134,  and the Observatoire de Paris and LERMA for
hospitality during this work. He also benefited from conversations
with  John P. Ralston.
\end{acknowledgments}

\end{document}